\documentclass[final]{ar-1col-S2O} % final draft

\usepackage[comma]{natbib}
\usepackage{url}

\jname{Annu. Rev. Astron. Astrophys.}
\jvol{63~9}
\jyear{2025}
\doi{10.1146/annurev-astro-021225-030604}

\usepackage{graphicx}
\graphicspath{{figs}}
\usepackage{color,soul,amsmath,amssymb, lscape,pdfpages} %% ulem
\usepackage[colorlinks=false,citecolor=blue,urlcolor=blue,breaklinks]{hyperref}

\def\aj{{AJ}}                   % Astronomical Journal
\def\araa{ARA\&A}             % Annual Review of Astron and Astrophys
\def\apj{{ApJ}}                 % Astrophysical Journal
\def\apjl{{ApJ}}                % Astrophysical Journal, Letters
\def\aap{{A\&A}}                % Astronomy and Astrophysics
\def\aapr{{A\&A~Rev.}}          % Astronomy and Astrophysics Reviews
\def\mnras{{MNRAS}}             % Monthly Notices of the RAS
\def\pasj{{PASJ}}               % Publications of the ASJ
\def\solphys{{Sol.~Phys.}}      % Solar Physics
\def\ssr{{Space~Sci.~Rev.}}     % Space Science Reviews
\def\nat{{Nature}}              % Nature
\def\jgr{{J.~Geophys.~Res.}}    % Journal of Geophysics Research
\def\planss{{Planet.~Space~Sci.}}           % Planetary Space Science
\def\pasa{Publications of the Astronomical Society of Australia}
\def\icarus{Icarus}

\let\mnrasl=\mnras

\begin{document}

% Page header
\markboth{Vidotto}{Star-Planet Interactions}

% Title
\title{Star--Planet Interactions: \\ A Computational View}

%Authors, affiliations address.
\author{A.~A.~Vidotto$^1$
\affil{$^1$Leiden Observatory, Leiden University, The Netherlands; \\email: vidotto@strw.leidenuniv.nl}}

%Abstract
\begin{abstract}
There are several physical processes that mediate the interaction between an exoplanet and its host star, with the four main ones being due to magnetic, particle (stellar outflow), radiative and tidal interactions. 
These interactions can be observed at different wavelengths, from X-ray to radio. Their strengths depend on  the architecture of planetary systems, as well as the age and activity level of the host stars. 
In particular, exoplanets in close-in orbits and/or orbiting active host stars can experience strong  physical interactions, some of which are negligible or absent in the present-day Solar System planets. 
Here, I present an overview of star-planet interactions through the lens of three-dimensional (3D) numerical models. The main conclusions are:

\hangindent=.3cm$\bullet$ Models are fundamental to interpret and  guide  observations. The powerful combination of observations and models allows us to extract important physical parameters of the system, such as, planetary magnetic fields, stellar wind properties, etc.

\hangindent=.3cm$\bullet$ The non-axisymmetric forces of the interactions generate spatially asymmetric features (e.g., planetary material trailing the orbit, shock formation), thus requiring the use of 3D models. 

\hangindent=.3cm$\bullet$ Star-planet interactions  vary in different timescales (from hours to giga-years) that are related to both planetary  (orbital motion, rotation) and stellar  (flares, cycles, and long-term evolution) properties. Understanding these variations  require time-dependent models. 

I advocate that future 3D models should be informed by multi-wavelength, (near-)simultaneous observations. The use of observations is twofold: some generate inputs for models (e.g., stellar magnetic field maps), whereas others are fitted by models (e.g., spectroscopic transits). {This combination of observations and models provides a powerful tool to derive physical properties of the system that would otherwise remain unknown.} 

\end{abstract}

%Keywords, etc.
\maketitle

%Table of Contents
\tableofcontents

%%%%%%%%%%%%%%%%%%%%%%%%%%%%%%
\section{MOTIVATION AND SCOPE OF THIS REVIEW} 
Most of the exoplanetary systems currently known have architectures that are very different from that of the Solar System. A striking difference is the existence of planets with very close orbits. The close proximities to their host stars imply that these planets are subjected to very strong star-planet interactions (SPIs). 

\begin{figure}[!h]
\includegraphics[width=0.99\textwidth]{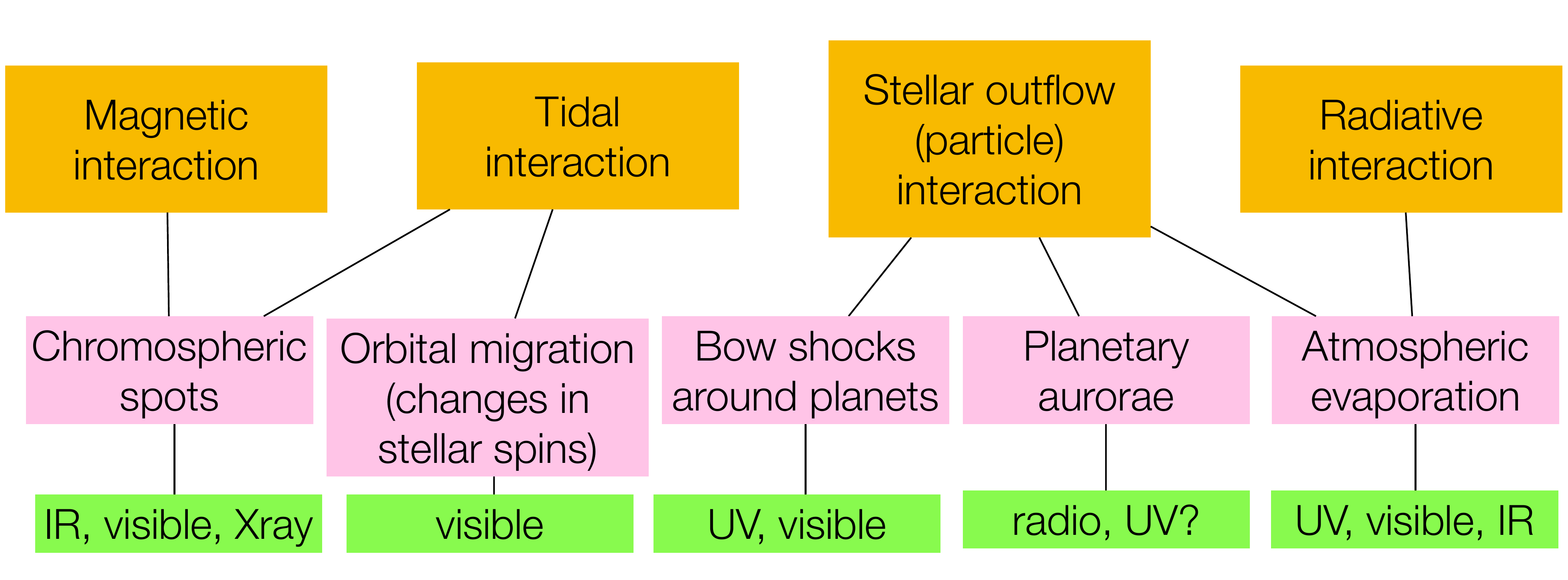}
\caption{Different types of star-planet interactions are shown in the top row (orange boxes). The middle row (pink boxes) shows some (non-exhaustive) physical phenomena that are associated with the several SPIs discussed in the text, whereas the bottom row (green boxes) shows possible wavelengths in which these phenomena can be observed. Adapted with permission from \citet{2020IAUS..354..259V}; copyright Cambridge University Press. Abbreviation: SPIs, star-planet interactions. \label{fig.spi_diagram}}
\end{figure}

SPIs can be classified into four different types, depending on the dominant forces involved  in the physical interaction: magnetic, tidal, particle (stellar outflows), and radiative interactions.  Figure \ref{fig.spi_diagram} shows a diagram of different types of interactions and some of the physical phenomena that they lead to. Figure \ref{fig.spi_diagram} also shows some of the possible wavelengths in which these phenomena can be observed. All planets, including those in the Solar System, are likely subjected to these four types of interactions during their lifetimes. However, the interaction strength, which is linked to the potential of SPI being detected in a system, varies from planet to planet and with time. Some extreme systems experience strong interactions originating from the four main processes. This is likely the case of the warm Neptune GJ436b, with observations indicating SPI of radiative \citep{2016A&A...591A.121B}, particle \citep{2023A&A...678A.152V}, magnetic \citep{2023AJ....165..146L} and tidal \citep{2021A&A...647A..40A} natures. For most of the planets, however, some of the interaction types are weak, and in most of the cases, only one or two SPI processes dominate in a given star-planet pair.   

\begin{marginnote} 
\entry{SPI}{Star-Planet Interaction.}
 \end{marginnote}

This article presents an overview of three-dimensional (3D) magnetohydrodynamics (MHD) numerical models developed in the study of SPIs. I focus   on interactions with close-in exoplanets, as these are more likely to show detectable signatures of SPIs. Of the four interactions illustrated in Figure \ref{fig.spi_diagram}, tidal interactions is the only SPI that is outside the scope of this article (as it is not a MHD process). This interaction is relevant for studying the evolution of architectures of planetary system, as it can lead to planetary migration, changes in stellar spins, as well as orbital circularisation, spin-orbit alignment and tidal heating \citep{2008CeMDA.101..171F, 2012A&A...544A.124B, 2016A&A...591A..45P, 2017CeMDA.129..509B, 2019ApJ...886...72M}. 
The remaining three interactions, namely magnetic, particle and radiative interactions, are tied to each other through stellar magnetism.
Outflows of low-mass stars, which are  the majority of host stars of known exoplanets, are magnetised and magnetically-driven, thus magnetic and particle interactions often cannot be separated, i.e., particle interactions do not exist without magnetic interaction and vice-versa.\footnote{In the literature, magnetic SPI is sometimes limited to  interactions where there is a magnetic coupling between the planet and the star, with stellar-outflow--planet interactions (which I refer to as particle SPI in this review) implicitly referring to interactions where such coupling does not take place. I discuss the conditions for star-planet magnetic coupling in Section \ref{sec.connectivity}.} 
Planets can also develop outflows, driven by the host-star's chromospheric/coronal high-energy radiation, which is ultimately generated by stellar magnetism. 
\begin{marginnote} 
 \entry{Low-mass stars}{Main-sequence stars with spectral types F, G, K and M. They are the majority of host-stars with detected exoplanets.}
 \end{marginnote}
\begin{marginnote} 
\entry{Transmission spectroscopy}{observation of stellar light as it is transmitted through the atmospheres of transiting planets; also referred to as spectroscopic transit. }
\end{marginnote} 

%%%%%%%%%%%%%%%%%%%%%%%%%%%%%%%%%%%%%%%%%%%%%%%%%%%%%%%%%%%%%%%%%
\section{SOME OBSERVED PHENOMENA ASSOCIATED WITH SPI}\label{sec.observations}

As illustrated in Figure \ref{fig.spi_diagram}, there are several physical phenomena that are thought to be associated with SPIs. Here, I present an overview of the historical developments of  three main physical phenomena associated with SPI: planet-induced activity on the host star,  atmospheric evaporation observed through transmission spectroscopy\footnote{Transmission spectroscopy is used to trace chemical elements present in the planetary atmosphere, as the spectral line probing the chemical species shows deeper transit depth than that seen in broadband optical observations.} and auroral emission at radio wavelengths. 
 
\subsection{Planet-induced stellar activity} Historically, several planet detection techniques, such as the transit method or radial velocity method, evolved from techniques formerly used in studies of binary stars. Likewise, seminal works of planet-induced activity due to SPIs followed the same route. Inspired by the high activity observed in some close-binary stars (whose activity can be much higher than in isolated stars),  seminal works put forward the idea that interactions with a close-in planet could induce  activity on their host stars. Initial suggestions proposed that magnetic reconnection between a close-in exoplanet and its host star could be the source of superflares \citep{2000ApJ...529.1031R} or cause hot spots through local chromospheric heating enhancement \citep[][see Supplemental Figure 1a]{2000ApJ...533L.151C}. 
Another proposed scenario,  also inspired by close-binary stars, was that tidal interactions could  generate local heating and thus formation of hot spots \citep[][see Supplemental Figure 1b]{2000ApJ...533L.151C}. 
In this case, tidal interactions between a host star and  a close-in planet would raise tidal bulges in the outer stellar atmospheric layers. With the planetary orbital motion, these bulges would then rise and subside over and over again, triggering waves and turbulent flows. Both  tidal and magnetic interactions could therefore affect the outer atmospheres of planet-hosting stars and generate non-radiative local heating that could be observed through an anomalous activity enhancement \citep{2000ApJ...533L.151C}.  Earlier theoretical models gave support to the generation of hot spots caused by magnetic reconnection events \citep[e.g.][]{2004ApJ...602L..53I, 2006MNRAS.367L...1M}. 
\begin{marginnote} 
\entry{Outer stellar atmospheric layers}{chromospheres, transition regions, and coronae}
\entry{Induced chromospheric hot spots}{Localised stellar spots induced by interactions with a close-in planet, that would generate  an anomalous increase in stellar activity.}
 \end{marginnote}

Identifying the type of interactions was originally done by monitoring of the anomalous emission and its periodicity \citep{2001MNRAS.325...55S}: while magnetic  interactions would  show periodicity modulated by  the orbital period $P_{\rm orb}$, tidal interactions would be modulated by $P_{\rm orb}/2$ (due to two tidal bulges). Today, we know that the periodicity of SPI signatures is likely more complicated, as they depend on the geometric properties of the system, stellar magnetic field configuration, as well as the  window of the observations \citep{2019ApJ...872..113F, 2023MNRAS.524.6267K}. However, it is still expected that some form of periodicity should exist.

\begin{figure}[t]
\includegraphics[width=0.99\textwidth]{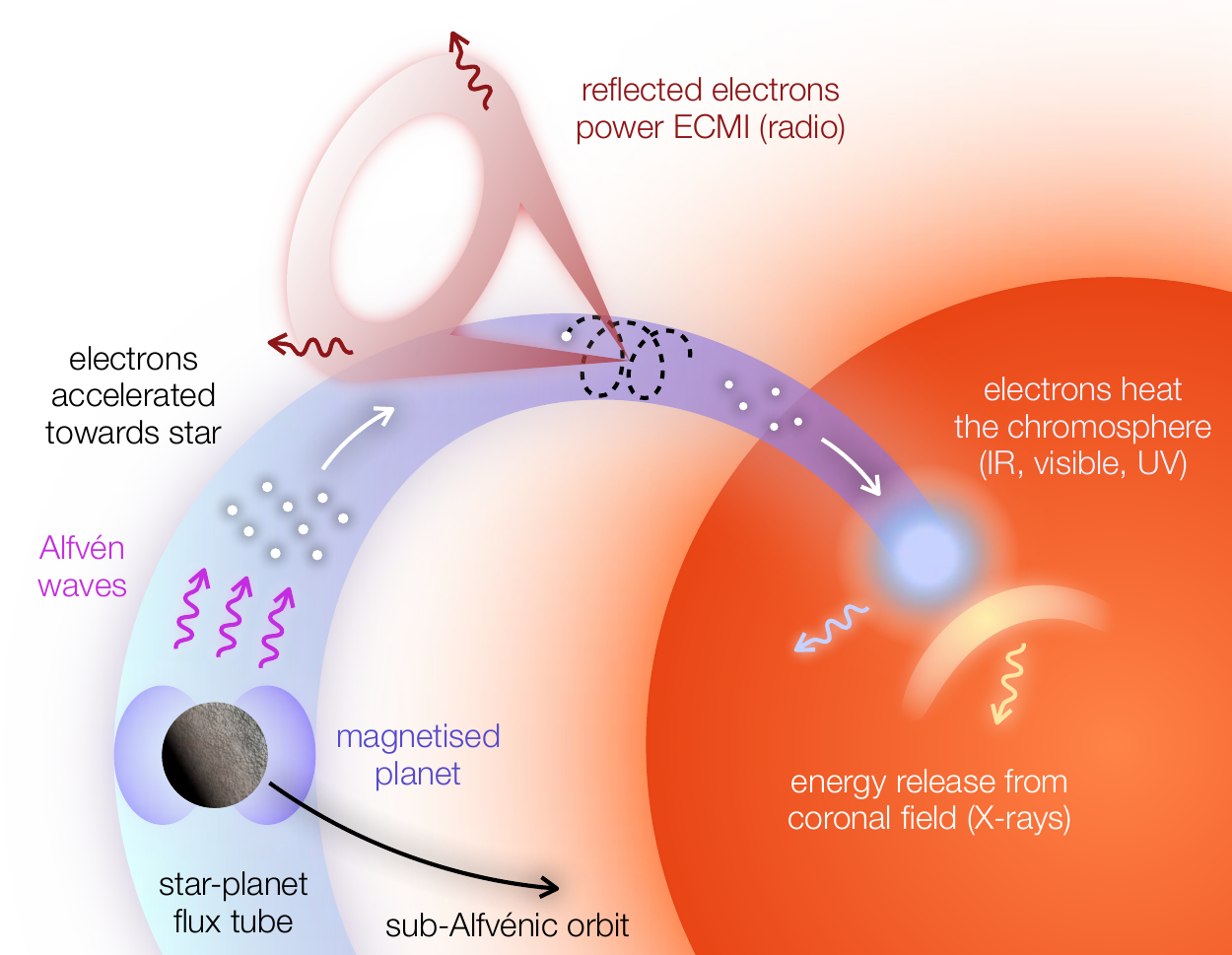}
\caption{Planet-induced activity in the host star can be probed at different wavelengths. This illustration shows a magnetised planet, but  planets without intrinsic magnetic fields can also induce activity in their host star. Figure reproduced with permission from \citet{2022PhDT.........8K}. Abbreviation: ECMI, electron-cyclotron maser instability. }\label{fig.lambda_kavanagh}
\end{figure}

The first theoretical works motivated numerous observational searches of activity enhancement in planet-hosting stars, with \citet{2003ApJ...597.1092S} presenting the first evidence of induced chromospheric hot spots associated with planet-induced activity in the exoplanetary system HD179949. In this work, they concluded that the anomalous activity observed in the chromospheric CaII K lines had its origin from magnetic SPI. Many other teams continued with observational studies and follow up monitoring of exoplanetary systems  with the goal of searching for activity enhancements caused by planets. A dedicated review of these works is beyond the scope of this article, but I would like to highlight two main findings. Firstly, the anomalous activity is not limited to chromospheric CaII K lines. Signatures can be generated in other wavelengths, such as alternative chromospheric lines like the HeI D3 line \citep{2022MNRAS.512.5067K}, coronal emission \citep{2011ApJ...741L..18P, 2015ApJ...805...52P}, and radio bursts \citep{2023NatAs...7..569P}, as illustrated in Figure \ref{fig.lambda_kavanagh}. Secondly, SPI signatures may not be repeatable. Follow-up observations of the anomalous activity observed in HD179949 showed that previously observed hot spots had disappeared \citep{2008ApJ...676..628S}. The non-repeatability of SPI signatures makes it hard to confirm and interpret detected signatures of SPI, sometimes casting doubt whether we are really observing SPIs or an artefact inherent of stellar variability itself \citep{2013A&A...552A...7S}. I will argue further in Section \ref{sec.future} that multi-wavelength near-simultaneous observations combined with modelling is the way forward to constrain SPI signatures and reduce potential controversy. In particular, knowledge of  stellar magnetic field topology and strength is fundamental in interpretations of magnetic SPIs.

\paragraph*{What can we learn from studies of Planet-induced activity?} We can derive planetary magnetic field strengths \citep{2005ApJ...622.1075S, 2019NatAs...3.1128C}. 

\subsection{Planetary evaporation measured through spectroscopic transits \label{sec.transmission}} Several years before the first detection of escaping planetary atmospheres, \citet{1998ASPC..134..241S} theoretically predicted that atmospheric material escaping close-in planets would form a comet-like  structure trailing the planet. In their seminal study, they argued that the escaping material would  reach altitudes beyond the Roche lobe, which would indicate that the  material was no longer gravitationally bound to the planet, and that it could be observed through spectroscopic transit observations. It was only a few years after this that the first detection of evaporating material escaping the atmosphere of the hot Jupiter HD209458b was reported by \citet{2003Natur.422..143V}, based on spectroscopic transits observed in the Ly-$\alpha$ line. The observations of the extended  H-rich atmosphere of HD209458b led to many observational studies of atmospheric escape in close-in planets (see \citealt{2023IAUS..370...56D} for a recent review). 

Hydrogen escape has historically been observed with the Hubble Space Telescope, using the Ly-$\alpha$ line in the ultraviolet band.  Ly-$\alpha$, however, has an important drawback -- the core of the line is contaminated by the interstellar medium and affected by geocoronal emission \citep{2019A&A...622A..46D}. As a result, only the wings of the line can be used to probe escape (radiative SPI). Additionally, the wings trace material at high velocities in the planetary upper atmosphere, which is also where the interaction with the stellar wind takes place (probing therefore particle SPIs).
Neutral helium in the near-infrared has also been used to probe atmospheric escape; this line does not suffer from contamination as in the Ly-$\alpha$ line, and can be accessed through ground-based observations \citep[e.g.,][]{2018Sci...362.1388N, 2020AJ....159..115K}. This evaporation diagnostics has been more intensively used recently, with the number of He escape detections now having surpassed that of Ly-$\alpha$ escape detection. Although most of the detections of atmospheric escape has been focused on light elements, heavier elements such as magnesium, carbon, silicon and iron have also been detected in escaping exoplanetary atmospheres \citep[e.g.,][]{2010ApJ...714L.222F, 2010ApJ...717.1291L, 2013A&A...551A..63B, 2023ApJ...954L..23S, 2024AJ....168..108E}. These observations open new avenues to characterise atmospheric evaporation, as they extend the range of spectral diagnostics used to probe exoplanetary mass loss. 

Observations alone do not provide stringent constraints on exoplanetary mass loss; quantities such as mass-loss rates, outflow velocities, densities and planetary magnetic fields remain unknown without modelling efforts. Therefore, in parallel to observational advances, a number of atmospheric escape models were developed. One of the first  models of exoplanetary atmospheric loss  relied on the energy-limit escape, which assumes that the energy flux deposited by high-energy stellar radiation goes into heating, which then drives the planetary outflow \citep{2003ApJ...598L.121L, 2004A&A...419L..13B, 2004A&A...418L...1L}. These models were then succeeded by more detailed hydrodynamical (HD) 1D models, which accounted for some major underlying physical processes, such as the chemistry, heating, cooling and tidal gravity \citep[e.g.,][]{2004Icar..170..167Y, 2005ApJ...621.1049T, 2007P&SS...55.1426G, 2009ApJ...693...23M}. More recently, multi-dimensional codes are employed to model planetary outflows  using fluid \citep{2007ApJ...671L..57S, 2011ApJ...728..152T, 2012ApJ...744...70K} and particle  \citep{2008Natur.451..970H, 2010ApJ...709..670E, 2013A&A...557A.124B} codes. Several of these studies demonstrated that key observables, such as absorption of stellar Ly-$\alpha$ emission, can only be properly assessed  with 3D simulations that accounts for the complex, asymmetric nature of planetary outflows \citep[e.g.,][]{2014MNRAS.438.1654V}, as is further discussed in Section \ref{sec.asymmetry}.

\paragraph*{What can we learn from spectroscopic transits?} We can derive atmospheric escape rate \citep{2007ApJ...671L..57S, 2013A&A...557A.124B, 2014A&A...565A.105B}; infer the local \citep{2013A&A...557A.124B, 2014Sci...346..981K} and global \citep{2017MNRAS.470.4026V, 2023A&A...678A.152V} conditions of the stellar wind (i.e., interplanetary medium); and estimate planetary magnetic fields \citep{2010ApJ...722L.168V, 2014Sci...346..981K}.

%%%%%%%%%%%%%%%%%%%%%%%%%%%%%%%%%%%%%%%%%%%%%%%%%%%%%%%%
\subsection{Auroral emission at radio wavelengths}  
A long-awaited SPI diagnostic is that of planetary auroral emission at radio wavelengths. This emission is observed in the magnetised planets in the Solar System \citep{1984Natur.310..755D} and, analogously, has been conjectured for exoplanets \citep{1999JGR...10414025F}. Such an emission is produced through the electron-cyclotron maser  (ECM) emission, and only exists if the emitting body is magnetised. The frequency of the emission takes place at the cyclotron frequency ($f_c [\textrm{MHz}] = 2.8 B_p  [\textrm{G}]$) and its first few harmonics \citep{1982ApJ...259..844M}, which implies that once we detect ECM emission from an exoplanet, there is a direct way for us to infer their magnetic fields $B_p$ from the observed frequency. Additionally, because the emission would be originating directly at the planetary magnetic field lines, detecting ECM emission could also be a way to directly detect exoplanets. 

\begin{marginnote} 
\entry{ECM}{Electron-cyclotron maser}
\entry{Radiometric Bode's law}{Empirical relation connecting the incident power of the solar wind and the output  radio power of the magnetised planets of the Solar System. }
\end{marginnote} 

For the Solar System objects, there exists an empirical relation, known as the radiometric Bode's law, in which the output radio power from the planet's auroral emission is proportional to the input power of the incident solar wind dissipated in the planetary magnetosphere  \citep{1984Natur.310..755D, 1998JGR...10320159Z}. This empirical relation has been extrapolated to close-in exoplanets, which are expected to generate radio emission substantially more powerful than that of the Solar System planets, which could, in principle, facilitate detection \citep[e.g.][]{1999JGR...10414025F}. This is because the input power from the stellar wind incident on close-in exoplanets is higher than that of the Solar System objects (e.g.,  higher ambient magnetic fields and ambient densities, \citealt{2015MNRAS.449.4117V}). So far, most the observational studies to detect exoplanetary radio emission have resulted in non detections. There are a number of reasons why this could be the case, including the planet being unmagnetised, the emission not oriented towards the observer,  the stellar wind being much weaker than predicted, or due to lack of instrument sensitivity. Recently, though, \citet{2021A&A...645A..59T} announced the potential detection of radio emission from the hot Jupiter $\tau$ Boo b  using LOFAR. Follow up monitoring to confirm or refute this detection continues to be conducted \citep{2024A&A...688A..66T}.
 
In terms of modelling auroral emission from exoplanets, initial studies have been done using semi-empirical prescriptions of stellar winds \citep{2005MNRAS.356.1053S, 2005A&A...437..717G, 2008A&A...490..843J} as a way to compute the input energy dissipated on the planet's magnetosphere.  It was only in the 2010's that the first models looking into the emission using 3D models of stellar winds started appearing \citep{2010ApJ...720.1262V}, including also models that consider the observationally-derived stellar magnetic maps  \citep{2015MNRAS.449.4117V, 2015MNRAS.450.4323S}. These models  have shown that the stellar wind conditions at the planetary orbit are not homogenous, which could affect the intensity of the emission. Regardless of their sophistication (i.e., 1D, 3D or semi-empirical), stellar wind models are needed to derive the winds' physical conditions at the planetary orbit: local wind density, velocity and magnetic fields. These quantities are then used to compute the kinetic and magnetic powers of the stellar wind. Finally, using the radiometric Bode's law, an output power and density flux for the radio planet are calculated, guiding therefore future radio observations. The predicted radio powers, however, all depend on the assumed planetary magnetic field strength \citep{2005A&A...437..717G} and topology \citep{2018A&A...616A.182V}, which are poorly constrained. 

\paragraph*{What can we learn from observations of radio auroral emission?} We can probe and characterise planetary magnetic fields \citep{2021A&A...645A..59T}; use radio observations as a potential exoplanet discovery technique \citep{2010AJ....139...96L};  and place constraints on stellar wind properties  \citep{2017A&A...602A..39V}. (A summary of the discussion from Section \ref{sec.observations} is shown in the sidebar titled ``What Can We Learn From SPI Studies?.'')

\begin{textbox}[h]%
\section{WHAT CAN WE LEARN FROM SPI STUDIES?}\label{greenbox_learn}%
SPI detections associated with modelling allow us to infer physical conditions of exoplanetary systems that would otherwise be very difficult to derive. This combined effort allows us to:
\begin{itemize}
\item probe the presence and strength of exoplanetary magnetic fields:  from auroral emission from exoplanets \citep{2021A&A...645A..59T};  from  auroral emission induced on the host star \citep{2023NatAs...7..569P}; from  anomalous chromospheric and coronal activity induced on the host star \citep{2005ApJ...622.1075S, 2019NatAs...3.1128C}; from bow-shock formation and detection \citep{2010ApJ...722L.168V};  from spectroscopic transits probing atmospheric evaporation  \citep{2014Sci...346..981K}. 
\item probe the stellar wind environment surrounding the planet, such as its density, stellar wind velocity and mass-loss rates \citep{2013A&A...557A.124B, 2014Sci...346..981K, 2017MNRAS.470.4026V, 2017A&A...602A..39V}.
\item derive escape rates of exoplanetary atmospheres \citep{2007ApJ...671L..57S, 2013A&A...557A.124B, 2014A&A...565A.105B}.
\item use radio observations as a novel exoplanet discovery technique in the context of radiometric Bode's law \citep{2010AJ....139...96L} and planet-induced stellar activity \citep{2020NatAs...4..577V}. 
\end{itemize} 
\end{textbox}

\section{WHY AND WHEN DO WE NEED 3D MODELS OF SPI? }\label{sec.asymmetry}
Over the years, most of the development in the field of SPIs came through  models with simplified symmetries, which have nevertheless provided important insights in the physics of SPI. For example, spherically symmetric (1D) models are excellent for incorporating more detailed chemistry in the hydrodynamics of escaping planetary atmospheres, which is numerically expensive to be treated in 3D. However, there are some limitations of what these models can predict, in particular considering that several physical processes involved in SPIs can remove the 1D symmetry of the system. I list some of these processes below:

\begin{itemize}
\item {\bf Stellar radiation} impacts the planet from the dayside. In the case of atmospheric escape driven by photoionisation from 
stellar X-ray and ultraviolet photons (XUV radiation), the incoming heating  produces a hotter dayside and a colder nightside. This day-night asymmetry changes the dynamics of the planetary outflow, removing the axi-symmetry of the system. In addition, planets that are tidally locked (likely to occur in close-in exoplanets) have the same hemisphere facing the host star, which can enhance even more the day-night asymmetry. 
\begin{marginnote} 
\entry{XUV radiation}{used to refer to radiation from X-ray plus ultraviolet wavelengths.}
\end{marginnote} 

\item {\bf The stellar wind} conditions around a planet are often inhomogeneous along the planetary orbit, due to the complexity of stellar magnetic fields \citep{2015MNRAS.449.4117V}. Additionally, because of high orbital velocities that close-in planets have, in the reference frame of the planet, the stellar wind impacts the planet at an angle with respect to the star-planet line \citep{2010ApJ...722L.168V}, thus having a substantial component along the orbital motion. The  interaction with the wind of the host star can generate a bow shock that partially surrounds the planet and its magnetosphere. In the case of close-in exoplanets, bow shocks  are more oriented towards the orbital motion \citep{2010ApJ...722L.168V}, as opposed to the case of Solar-System planets, where bow shocks form more or less in the dayside of the planet. As seen in the sketch in Supplemental Figure 2, there is therefore a mismatch between the orientation of the incoming stellar radiation (i.e., from the dayside) and the wind interaction region (i.e., having a component along the orbital motion), breaking the symmetry of the system. 

\item {\bf Planetary outflows  from their escaping atmospheres} are bounded by the winds of their host stars. Combined with the orbital motion, the stellar wind can sculpt escaping atmospheres and lead to the generation of a comet-like tail structure trailing the planet. In addition to sculpting the evaporating material, the stellar wind pressure exerted on the  expanding planetary outflow can also modify the exoplanetary mass loss, including the amount of escaping material, their densities and velocities \citep{2011ApJ...730...27A, 2011ApJ...728..152T, 2020MNRAS.494.2417V,2020MNRAS.498L..53C}. Together, these two effects generate an  asymmetric distribution of material around the planet, which leads to asymmetric spectroscopic transits: line profiles are not symmetric with respect to line centre, and transit lightcurves are also not symmetric with respect to mid-transit (see the sidebar titled Planetary Winds or Outflows?).

\begin{textbox}[h]% 
\section{PLANETARY WINDS OR OUTFLOWS?} 
When talking about planetary outflows, I avoid using the term `planetary wind', as it can be interpreted in the geophysical sense (East-West winds, weather and climate). Here, instead, I prefer using the terminology `planetary outflow' to remove ambiguity. This outflow from the planet takes a similar meaning to the Astrophysical `stellar wind', i.e., that of a bulk outflowing gas escaping the central object (planet or star). In this review, I have replaced `planetary wind' with `planetary outflow' and, instead of adopting the terminology of `wind-wind interaction', I use `flow-flow interaction'. Flow-flow interactions belong to the particle SPI classification shown in Figure \ref{fig.spi_diagram}.
\end{textbox}

\item {\bf Planetary magnetic fields} can also break the symmetry in SPIs. Magnetic fields change the dynamics of escaping atmospheres \citep[e.g.,][]{2015ApJ...813...50K,2021MNRAS.508.6001C} and their effects in a fluid-like escaping atmosphere resembles very closely that of magnetised stellar winds \citep[e.g.][]{2009ApJ...699..441V} -- an intrinsic dipolar field, for example, can create dead zones that do not participate in the mass-loss process  \citep{2011ApJ...730...27A, 2011ApJ...728..152T, 2015ApJ...813...50K,  2021MNRAS.508.6001C}. Finally, depending on the magnetic coupling between the exoplanet and the host star, planetary magnetic field lines can be connected to the host star \citep{2015ApJ...815..111S}. I will discuss more about magnetic coupling in Section \ref{sec.connectivity} (see also the sidebar titled Magnetism in SPI).    

%%%%%%%%%%%%%%%%%%%%%%%%%%%%%%%%%%%%%%%%%%%%%%%%%%%%%%%%%%%%%%%%%
\begin{textbox}[!h]%
\section{MAGNETISM IN SPI}
Magnetism is involved in all the SPIs discussed in this article. Radiative SPI  leading to photoevaporation of planetary atmosphere  takes place in planets orbiting low-mass stars that are magnetised, as their chromosphere, corona and associated high-energy radiation mostly originate from stellar magnetism \citep{2003ApJ...598.1387P, 2014MNRAS.441.2361V}. Particle SPI that are caused by stellar winds and  coronal mass ejections (CMEs) have also their origin in stellar magnetism, as outflows from low-mass stars are magnetically driven \citep{2013PASJ...65...98S}. Finally, both stellar and planetary magnetism directly participate in magnetic SPIs. In the case of planets, a relatively strong magnetic dipolar field creates a bubble-like structure around the planet, shielding at least part of the planetary surface against stellar outflows and cosmic particles. These ionised particles are deflected around  planetary magnetic field lines, similar to the solar wind--Earth's magnetosphere interaction. 

\subsubsection{The Strength of Planetary Magnetic Fields}\\

The combined effort of observing and modelling SPIs is likely the main avenue for uncovering exoplanetary magnetic fields.  There have been many efforts to quantify exoplanetary magnetic fields through a range of techniques \citep[e.g.,][]{2008ApJ...676..628S, 2010ApJ...722L.168V, 2015ApJ...810...13C, 2019NatAs...3.1128C, 2021A&A...645A..59T}. These studies found magnetic field strengths in the range of 10 -- 120~G for close-in gas giants, and theoretical studies can even predict higher field strengths. Recent dynamo studies  indicate that the field strength does not grow indefinitely with rotation, but instead is determined by the energy flux emerging from the dynamo region \citep[e.g.,][]{2009Natur.457..167C, 2011Icar..213...12D}. Factors like the mass of the planet, its orbital distance, internal structure and age affect the dynamo-generated field in exoplanets \citep{2010A&A...522A..13R,2017ApJ...849L..12Y, 2021ApJ...908...77H, 2024MNRAS.535.3646K}.

\subsubsection{Stellar Magnetism -- Observations and Reconstruction}\\

Stellar magnetism can be indirectly inferred through activity diagnostics, such as those caused by photometric variations due to spots and plages, chromospheric and coronal emission in spectral lines, flares, etc. However, the  way to directly diagnose stellar magnetic field is via the observational signatures in stellar spectral lines of the Zeeman effect  \citep{2021A&ARv..29....1K}. Two techniques have provided most of stellar magnetic field measurements we currently have. 1) The Zeeman broadening technique uses unpolarised light to derive unsigned surface-average magnetic fields that are organised both in small and large scales \citep[i.e., the integrated field][]{2012LRSP....9....1R}. This technique does not constrain the sign (i.e., the geometry) of the stellar field. 2) The Zeeman Doppler imaging (ZDI, \citealt{1989A&A...225..456S}) technique provides the geometry of the surface large-scale magnetic field. Together, these techniques have provided magnetic field measurements of several hundreds of cool stars \citep{2009ARA&A..47..333D, 2021A&ARv..29....1K, 2022A&A...662A..41R}. These measurements are essential inputs for characterising SPIs. In particular, the large-scale field derived in ZDI studies has been used in 3D models of stellar winds  \citep[e.g.,][]{2015MNRAS.449.4117V} or 3D magnetic field extrapolations \citep[e.g.,][]{2010MNRAS.406..409F}, both of which allow us to characterise the magnetic environment around exoplanets and the strengths of SPIs.
 \end{textbox}

\item {\bf A planet's own rotation, orbital motion, radiation pressure and charge-exchange} are additional physical processes that remove symmetry. Orbital motion and stellar tidal gravity  lead to non-axisymmetric accelerations in the reference frame of the planet. Additionally, the orbital spin axis of exoplanets might not be aligned with the planetary rotation spin axis, also contributing to non-axisymmetric forces. Radiation pressure, similar to any incoming stellar radiation,  is oriented radially away from the star and reaches the planet from the dayside. Finally, charge-exchange processes can occur between the impacting ionised stellar wind and the neutral escaping planetary material. These processes preferentially take place in the interface of the flow-flow interaction, such as in the shock surrounding the planet, as well as in the flanks of the comet-like tails \citep{2019MNRAS.487.5788E}. 

\end{itemize}

%%%%%%%%%%%%%%%%%%%%%%%%%%%%%%%%%%%
\section{FUNDAMENTALS OF 3D MODELS FOR STUDYING SPIs}
In this section, I present some background material regarding the fundamentals of 3D models used for studying SPIs. I expand on some commonly found modelling setups, discuss the physical differences and similarities between outflows from exoplanets and from host stars, and also present the concept of star-planet magnetic coupling.

%%%%%%%%%%%%%%%%%%%%%
\subsection{Commonly found modelling setups} 
There is a variety of modelling setups used in simulations of SPIs. The definitions adopted in this review article are summarised below.

\subsubsection{Local versus Global simulations} 
\begin{figure}[!h]
\includegraphics[width=1.2\textwidth]{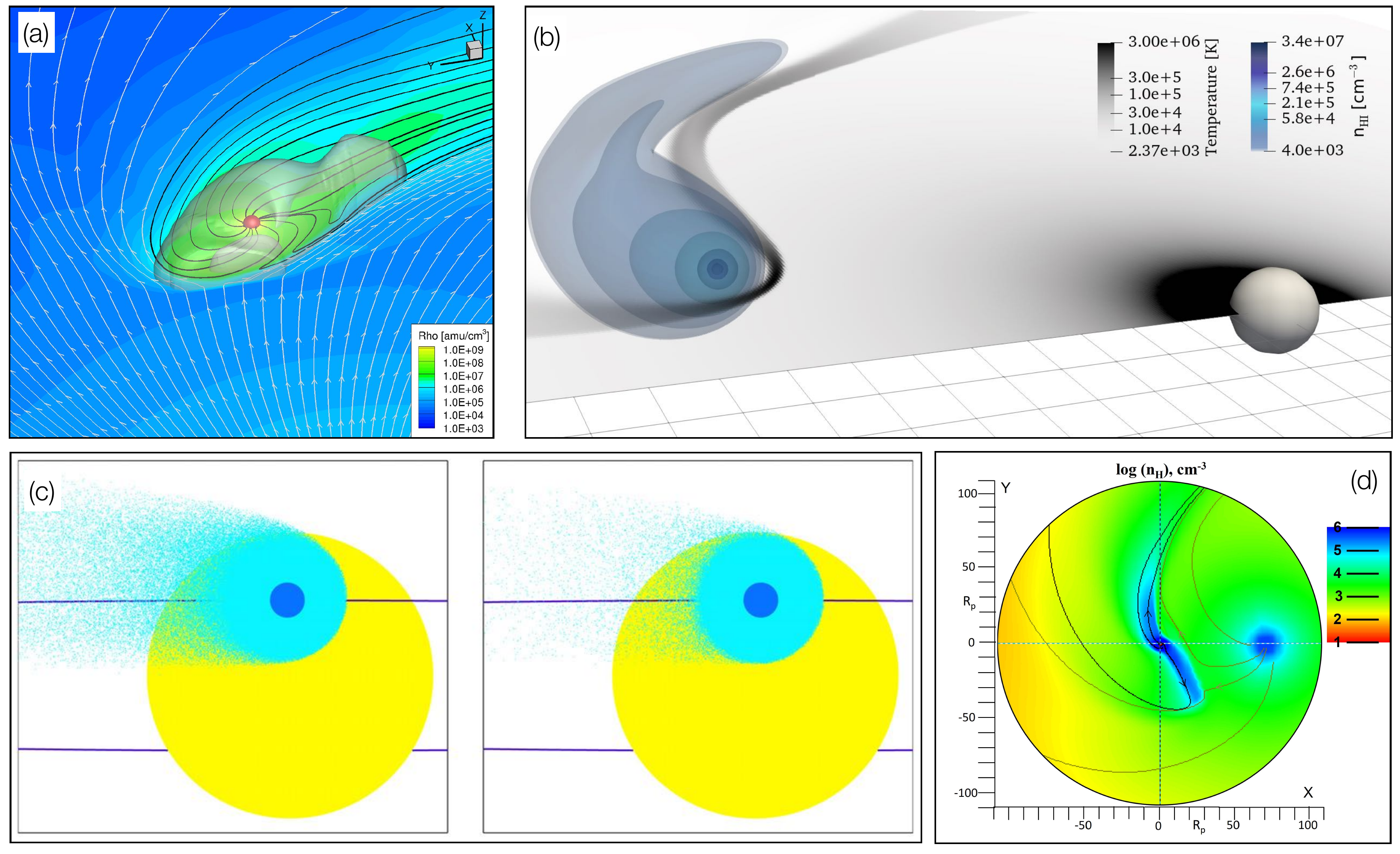}
\caption{Simulations of escaping atmospheres of close-in gas giant planets using different modelling setups. (a) Local, hydrodynamical simulation of the planetary outflow (black streamlines) interacting with a stellar wind (grey streamlines) that is injected in the numerical domain from $-x$.  (b) Global, hydrodynamical model of the host star GJ 436  (sphere on the right) and its orbiting planet (sphere on the left), whose escaping atmosphere form a comet-like tail structure.
(c) Particle simulations  of the escaping atmosphere of HD209458b (blue circle)  considering two different values of ionising fluxes: the light blue dots represent neutral hydrogen escaping the planet. (d) Multi-fluid simulations showing the proton density and its velocity streams. The planet WASP-80b is at the origin and the  star is at $x=76$ planetary radii.
Panel $a$  adapted from \citet{2021MNRAS.500.3382C}; copyright 2020 The Author(s). Panel $b$ reproduced from \citet{2021MNRAS.501.4383V}; copyright 2020 The Author(s). Panel $c$ adapted from \citet{2013A&A...557A.124B};  copyright ESO. Panel $d$ adapted from \citet[][CC BY 4.0]{2023A&A...673A..37F}. 
}\label{fig.local_global} 
\end{figure}

Local simulations consist of simulations where the host star is not included within the simulation domain. In this setup, the inner boundary is placed at or close to the planet's surface, and the effects of the star, such as its incoming radiation and winds, are included in the simulation domain through outer boundary conditions. Figure \ref{fig.local_global}a illustrates one such simulation, where we see the planet centred at the origin of the grid, with the stellar wind flow being injected from the $-x$ direction (bottom part of the image). By contrast, global simulations,  consider both the planet and the star directly in the numerical domain, and each body have their own set of boundary conditions. Naturally, this modelling setup covers larger physical extensions. Figure \ref{fig.local_global}b illustrates one such simulation, where we see the star on the {right} portion of the grid, interacting with the evaporating planet on the {left} portion of the grid. 

Local and global modelling setups have their advantages and drawbacks. One important advantage of global models, which is a disadvantage of  local simulations, is that, if the planet has any effect on the star, this effect can be captured in the models. For example, planetary material that can be accreted onto the star might create stellar chromospheric hot spots (Section \ref{sec.observations}). Local models cannot track this material once it leaves the simulation domain; in contrast, global models can track the material back to the star.
In general, the drawback of global setups is that they need more computational resources to provide the same numerical resolution as the local simulations offer. An alternative to make global simulations less computationally expensive is to simplify the equations that are being solved. For example, some global simulations of SPIs do not include the planetary outflow or include the acceleration of the outflow in a more simplified way, such as, without accounting for photo-ionisation driving \citep[e.g.,][]{2015A&A...578A...6M, 2015ApJ...815..111S,2018MNRAS.479.3115V, 2021MNRAS.500.3382C, 2024A&A...683A.226C}.

%%%%%%%%%%%%%%%%%%%%%
\subsubsection{Fluid versus Particle models:} Interactions between stellar winds and planetary outflows have been modelled using fluid or particle models. In fluid models, it is implicitly assumed that the gas is collisional and, thus, it obeys the (M)HD equations. In this case, only the macroscopically-averaged properties of the different fluids (such as, temperature, bulk velocities, density)  are computed in the model. The models depicted in Figure \ref{fig.local_global}a,b fall in the category of fluid models. In contrast, in particle models, a certain number of atoms are grouped together to form meta-particles, and collisions between particles are included explicitly in the numerical algorithm. In this type of model, each meta-particle is tracked during computation. Figure \ref{fig.local_global}c illustrates  such model. Some alternative models also combine  fluid and particle treatments, giving rise to hybrid models \citep{2018JGRA..123.1678J, 2019MNRAS.488.2108E}. The scope of this review limits our discussion to  HD and MHD models (i.e., fluid-type models), but it is worth recalling the importance of particle models in the literature of planetary evaporation \citep{2008Natur.451..970H, 2010ApJ...709..670E, 2013A&A...557A.124B}. One drawback of particle models is that the denser the environment is, the more particles they have to trace and thus the more computationally expensive they become. Although particle models can  be used to model collisional gas, fluid models might be more computationally advantageous in these cases. In the case of close-in planets, whose atmospheric temperatures are sufficiently high to create expanded, high-density atmospheres, the gas is collisional and can be treated as a fluid. The majority of models currently used to treat evaporating atmospheres of close-in planets rely on the fluid formalism (see the side bar titled Key Requirement For Hydrodynamic Escape of Planetary Atmospheres).

\begin{textbox}[h]%
\section{KEY REQUIREMENT FOR HYDRODYNAMIC ESCAPE OF PLANETARY ATMOSPHERES}
Hydrodynamic escape only works in the limit that the planetary atmosphere can be treated as a fluid. This means that the atmosphere needs to be collisional. This condition can be indicated by the Knudsen number 
$$\textrm{Kn}=\frac{\lambda_{\rm mfp}}{H} > 1 \, ,$$
where ${\lambda_{\rm mfp}}$ is the mean free path of an atmospheric particle to Coulomb collisions and $H$ is the atmospheric scale height. Hydrodynamic codes usually check if this condition is met after the hydrodynamic escape solution is found. It is sufficient for $\textrm{Kn}>1$ within the sub-sonic regime of the escaping atmosphere \citep{2009ApJ...693...23M}. If the atmosphere cannot be treated as a fluid, then a particle treatment to calculate evaporation rates and also trajectories of each (meta)particle is required.

Either the particle or fluid treatments provide velocity, density and temperature distributions that can be used to compute observables, such as, line profiles of spectroscopic transits.
\end{textbox}

%%%%%%%%%%%%%%%%%%%%%
\subsubsection{Single-fluid versus Multi-fluid simulations:} Among the  HD and MHD models, there is also a difference on how each species is treated. For example, models can treat neutral and ionised hydrogen as if they were two different species that undergo enough collisions with each other, such that they behave as a single fluid. These types of codes are known as single-fluid models with multi-species treatment. In this case, there is one set of HD equations (due to mass, momentum and energy conservations) that describe the different species combined, so that they all have a common bulk velocity,  pressure and  temperature. In the case that one wishes to trace the ionisation state of hydrogen, for example, (i.e., a multi-species treatment) one additional equation is needed, namely, the ionisation balance equation. Models depicted in Figures \ref{fig.local_global}a,b are examples of such setups.
 In contrast, in multi-fluid simulations, each  species is described by their own fluid properties, which means one needs to solve one set of HD equations for each species, such that each species will have their own velocity, temperature and pressure distributions. These different fluid-species then interact with each other  through collisions. In multi-fluid models, heating and cooling processes can be computed in an additional equation for the electron pressure. The energy of the electrons is then transferred to the other fluids via collisions, but because of their much lower masses, their momentum is usually neglected \citep{2009JGRA..11412203G, 2012JCoPh.231..870T}.
 
 These different formalisms have advantages and disadvantages.  One  disadvantage of multi-fluid models is that they are more computationally expensive than single-fluid multi-species models: for each fluid-species considered, a new set of equations (mass, momentum and energy) need to be computed, resulting in five additional equations in 3D models to describe the HD properties of each added fluid. However, there are certain physical conditions that can only be properly accounted for with multi-fluid formalism. For example,  when collisions between different fluid-species are less frequent,  this can lead to a decoupling between light and heavy species, resulting in fractionation of elements  \citep{2023ApJ...953..166X}. Because single-fluid treatments do not discriminate between the velocities of each species, the abundance of species remains constant with height in the atmosphere. In a multi-fluid treatment, it is possible to self-consistently compute how much the hydrogen flow can actually drag other species via collisions, leading to a departure of constant abundance throughout the atmosphere.

\subsection{Differences and similarities between stellar and planetary outflows}
Planetary outflows that meet the fluid condition behave  similar to stellar outflowing winds. Both the stellar and the planetary outflows start with a nearly zero velocity and accelerate as they propagate outward. Provided there is enough space for the full acceleration to take place, at a certain distance, the outflow reaches asymptotic speed, known as terminal velocity. Because hydrodynamical outflows are continuously accelerated, they do not need to be above surface escape speed to be escaping.
\begin{marginnote} 
\entry{Surface escape speed}{Speed required for a particle to escape the gravitational potential of an object. }
\entry{Terminal velocity}{Asymptotic speed reached by an outflow. }
\end{marginnote} 

\subsubsection{Critical radii in HD and MHD solutions}
Similar to stellar wind theory, the physical solution for planetary outflows has critical radii that determine outﬂow properties, such as its mass-loss rate \citep{1958ApJ...128..664P, 1967ApJ...148..217W}. In the case of non-magnetised flows,  the critical radius is known as the sonic radius, which is the radius where the velocity of the escaping  outﬂow reaches sonic speed. In the case of magnetised flows, an additional critical radius is the Alfven radius, which is located at the position where the velocity of the outﬂow reaches the Alfven speed. {In a 1D geometry, the critical radius is a point, whereas in 3D geometries, the critical radius takes the form of a surface.}\footnote{{Recently, solar wind observations and theory are indicating that the Alfven surface could instead be a fractal zone \citep{2023SoPh..298..126C}. In the discussion through this manuscript, I consider only the traditional view of a smooth, albeit irregular critical surface (as seen in Section \ref{sec.connectivity}).}} 

In spite of the qualitative similarities in the velocity structure between stellar and planetary outflows, quantitatively, the outflow velocities are quite different. For close-in gas giants, flow speeds are of the order of tens of km/s, whereas for solar-like stars, flow speeds are of several hundred of km/s or more. The reason for this difference is that planetary and stellar outflows are driven by different mechanisms. Winds of low-mass stars are thought to be heated and accelerated by MHD waves \citep[e.g.,][]{2008ASPC..384..317C,2013PASJ...65...98S}. By contrast, outflows from close-in exoplanets are primarily heated by incoming stellar XUV irradiation  \citep[e.g.,][]{2007P&SS...55.1426G}, with possible contributions of thermal energy, released after formation, and gravitational energy, released from contraction, \citep[e.g.,][]{2024MNRAS.528.1615O}. For both objects, the added heating from waves (stars) or photoionisation (planets) leads to a temperature increase and a pressure gradient force that contributes to the acceleration of the flow. Other forces involved in the momentum equation are the object's own gravitational force, tidal and Coriolis forces, and magnetocentrifugal acceleration (see the sidebar titled Planetary Outflows Escape With Speeds Below Surface Escape Speed).

\begin{textbox}[t]% 
\section{PLANETARY OUTFLOWS ESCAPE WITH SPEEDS BELOW SURFACE ESCAPE SPEED}
Because stellar winds and planetary outflows continue to be accelerated after they are ejected from the central object, their initial ejection speeds are very small and even below the surface escape speed.  In the literature of planetary escaping atmospheres, there has been a recurring misunderstanding, namely that if the detected outflow velocities are below surface escape speeds, authors assume atmospheres are not escaping. This is true under the Jean's escape mechanism, but it is incorrect in the case of photo-evaporation-driven outflows (hydrodynamic escape). The solar wind is a straightforward counter-example of such misunderstanding: the Sun has a surface escape speed of around 600 km/s, whereas its terminal wind speed is around 400 km/s.
\end{textbox}

\subsubsection{Sub- and super-critical disturbances}\label{sec.subcritical}
Disturbances propagating in fluids can travel at different speeds depending on the thermal and magnetic properties of the plasma \citep[e.g.][]{Priest_2014}. Some simple wave modes  commonly discussed in Astrophysical plasmas are: the sound wave traveling at the sound speed $c_s$, the Alfven wave travelling at the Alfven speed $v_A$, and the fast magnetoacoustic wave that propagates with a velocity of $(v_A^2 + c_s^2)^{1/2}$ (assuming propagation perpendicular to the magnetic field). If a disturbance is generated where the fluid velocity exceeds its propagation speed, then the disturbance cannot propagate oppositely to the flow and will instead be carried out by the outflow. For example, disturbances generated beyond the sonic radius will be carried away by a sound wave. In contrast, disturbances generated below the sonic radius can be carried back to the star/planet by sound waves. This example can be extended to other types of waves and their respective critical radii \citep[such as the Alfven radius,][]{1967ApJ...148..217W}.

\begin{figure}[t]
	\includegraphics[width=1.2\textwidth]{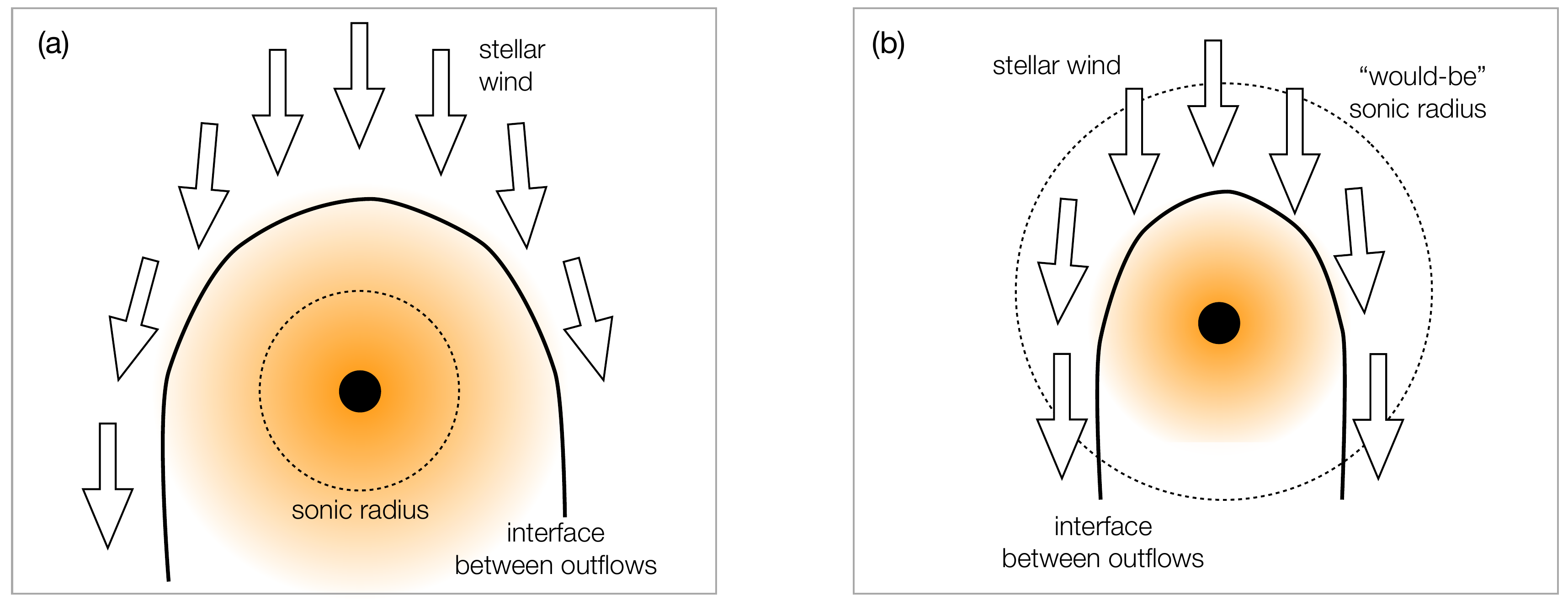}
	\caption{Sketch of how stellar outflows (arrows) can affect the structure of escaping planetary atmospheres (orange region), depending on whether the critical radius (dashed) lies inside (a) or outside (b) the interacting boundary between stellar and planetary outflows (thick black curve). Figure adapted from \citet{2020MNRAS.494.2417V}; copyright 2020 The Author(s). \label{fig.pl_conditions} } 
\end{figure}

The propagation of waves is very important when considering how the outflow responds to external disturbances. For example, the position of the interface between stellar and planetary outflows with respect to the critical radii of the planetary outflow (i.e., the sonic radius, or Alfven radius, depending on whether the planet is magnetised or not) is key for determining whether exoplanetary mass loss is affected by the stellar wind   \citep{2016ApJ...820....3C, 2020MNRAS.494.2417V, 2020MNRAS.498L..53C}. Planetary outflows are bounded by the winds of their host stars, which exert a significant, non-zero pressure on what would otherwise be a freely expanding flow; the external pressure confinement  is particularly important for close-in planets and/or planets orbiting more active stars. Consider these two conditions, as illustrated in Figure \ref{fig.pl_conditions}. In the first case (panel a), the critical  radius  (e.g., sonic or Alfvenic) of the planetary outflow is inside the interface that separates star-planet outflows. Here, the inner structure of the planetary outflow  cannot be affected by the stellar wind. This is because the super-critical velocity of the outflow carries away any disturbances and they do not propagate back to the planet. In this case, the stellar outflow can only delineate/shape the outer part of the escaping atmosphere -- it affects the signature that could be detected with transmission spectroscopy (e.g., asymmetric line profiles, velocities), but does not change the inner structure of the escaping atmosphere nor its escape rate. In the second case (panel b), the  interface separating star-planet outflows is deep enough so that the stellar wind interacts with the still sub-critical (e.g. sub-sonic, sub-Alfvenic) planetary outflow. Here, because the disturbance is generated below the critical radius, it can be carried back to the planet. In this case, the atmospheric structure is altered by the stellar wind and models have shown that there is a suppression or reduction in escape rate \citep{2016ApJ...820....3C, 2020MNRAS.498L..53C, 2021MNRAS.500.3382C, 2024MNRAS.534.3622P}. 

\begin{figure}[t]
	\includegraphics[width=1.2\textwidth]{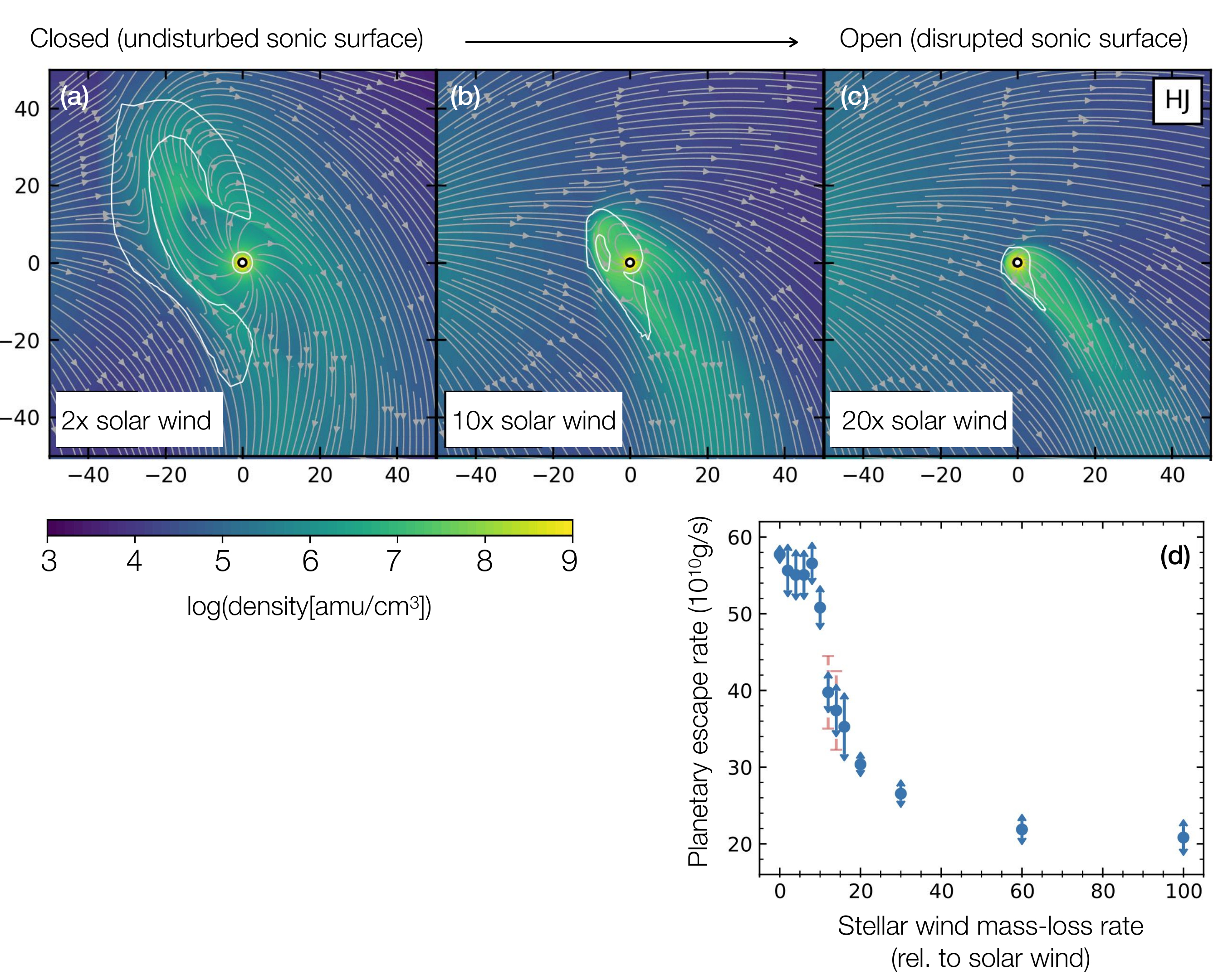}
	\caption{The effects of the stellar wind external disturbance on planetary atmospheric escape rates. Panels (a)-(c) show the interaction region between the stellar wind (injected from the left) and the planetary outflow, where the circle at the origin is the planet and the white contours show the sonic surface, where the velocity of the flow (stellar or planetary) reach sound speed. Panel (d) shows that the planetary escape rate is initially insensitive to the stellar wind conditions, until the stellar wind disrupts the sonic surface of the planetary outflow. Beyond this point (at around 10 times the solar wind mass-loss rate), the escape rate of the planet suffers a reduction. Figure adapted from \citet{2021MNRAS.500.3382C}; copyright 2020 The Author(s). \label{fig.carolan21}} 
\end{figure}

Figure \ref{fig.carolan21} shows the results of 3D HD simulations, where we see how the planetary escape rate responds to the injected stellar wind properties. The stellar wind is injected from the left part of the grid as seen in these equatorial cuts (panels a to c). As the wind mass-loss rate increases, the interaction region between the stellar wind and planetary outflow moves closer to the planet. This is because, as one increases the stellar wind density, its ram pressure gets larger and thus the balance point moves closer to the planet. At the same time, the original sonic surface of the planetary outflow (seen as a small white circumference surrounding the planet in panel a) is disrupted as the stellar wind mass-loss rate increases from 2 (panel a) to 10 times the solar wind values (panel b). For the additional scenario plotted in the  panel c, we see that the original sonic surface has been completely modified. In Figure \ref{fig.carolan21}d, we see the quantitative effects on the evaporation rate of the planet: as the stellar wind mass-loss rate increases, initially, the planetary escape rate remains approximately constant at around $55 \times 10^{10}$g/s. When the sonic surface of the planet starts to be disrupted, at about 10 times the solar wind value, we see the escape rate starts to be affected. In this series of models, the effects are reductions in escape rate by a factor of 3 from increasing   the stellar wind mass-loss rate by two orders of magnitude. These higher mass-loss rates are expected in younger solar-like stars \citep{2021LRSP...18....3V}. 

A similar concept can also be adopted regarding the position of the planetary orbit with respect to the critical radii of the stellar wind. This is an important point  to determine the types of SPIs of a given system, as I will present next.

%%%%%%%%%%%%%%%%%%%%%%%%%%%%%%%%%%%%%%%%%%%%%%%%%%%%%%%%%%%%%%%%%
\subsection{Star-planet magnetic coupling (connectivity)} \label{sec.connectivity}

The type of SPI and also their observational diagnostics depend on whether the interaction between the exoplanet and the stellar wind is super- or sub-Alfvenic. This difference is also relevant for deciding on the most appropriate numerical setup for a given study. In practice, the Alfven surface delineates the regime of the stellar wind that is dominated by magnetic forces (within the surface) or by inertial forces (beyond the surface). Inside the Alfven surface, the wind velocity is sub-Alfvenic, and information and energy can be transported from the planet to star through Alfven waves (e.g., \citealt{2013A&A...552A.119S}).\footnote{To be more precise, this refers to the stellar wind velocity in the reference frame of the planet, i.e., considering also orbital motion.} However, in the super-Alfvenic regime, this direct magnetic connection ceases to exist, and dynamical structures from the interaction, such as a bow shock, develops surrounding the planet.   

\begin{marginnote}
\entry{Magnetic coupling} {refers to a magnetic connection between star and planets in sub-Alfvenic motion through the stellar wind.}
\end{marginnote}

\begin{figure}[t]
	\includegraphics[width=1.2\textwidth]{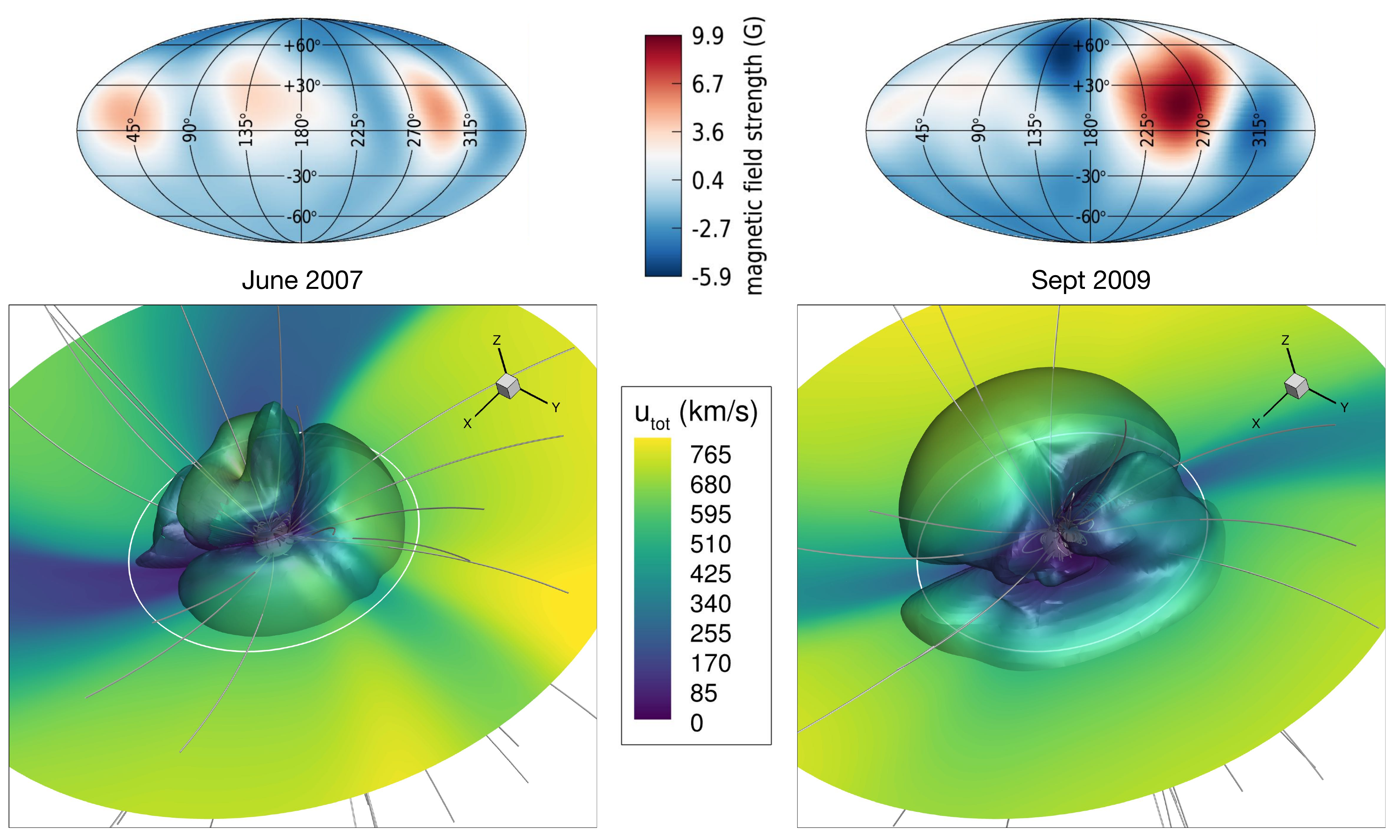}%
	\caption{\it (a) The radial component of the large-scale magnetic field of the planet-hosting star HD179949 in June 2007 and Sept 2009. These maps were reconstructed in  \citet{2012MNRAS.423.1006F} using Zeeman Doppler imaging. (b) 3D simulations of the wind of HD179949. These simulations use the maps shown in the top row as  boundary conditions. Here, the stellar wind speed is shown with the 
colourmap, the orbit of the hot Jupiter HD179949b is represented by the white circumference at 7.9$R_\star$, the Alfven surface is the irregular surface plotted in each of the bottom panels, and the grey lines show stellar magnetic field lines that are embedded in the stellar wind. \label{fig.salfven} } 
\end{figure}

To illustrate this, Figure \ref{fig.salfven}a shows the stellar magnetic field maps  of the planet-hosting star HD179949 reconstructed using ZDI \citep{2012MNRAS.423.1006F} at two different  dates: June 2007 (left) and Sept.~2009 (right). Figure \ref{fig.salfven}b shows the results of the  stellar wind  simulations, which use the corresponding magnetic maps as an inner boundary condition. This condition is the only difference between the two simulations, the remaining  parameters of the model are kept the same\footnote{These parameters are the same as those used in \citet{2023A&A...678A.152V} for Model II, which are the values usually adopted in solar wind simulations \citep[e.g.,][]{2014ApJ...782...81V}.}. Figure \ref{fig.salfven}b shows the orbit of the hot Jupiter HD179949b and the Alfven surface is the irregular surface shown in each subpanel. In the case of June 2007, the planetary orbit is always super-Alfvenic, indicating that there is no magnetic coupling between the star and the planet. By contrast, in the case of Sept.~2009, we see that most of the planetary orbit lies within the Alfven surface. At this epoch, magnetic coupling between the star and the planet can develop.

This study illustrates two challenges in the observations of SPI: first, it shows that any SPI signature would have evolved from June 20007 to Sept.~2009 (see Section \ref{sec.timescales}), and  could even have potentially disappeared \citep[e.g.,][]{2008ApJ...676..628S}. Second, SPI signatures might originate from different sources depending on the epoch of the observation. For example, at radio wavelengths, ECM emission can originate from electrons spiralling along the planetary and/or stellar magnetic field lines. In June 2007, when HD179949b is in a super-Alfvenic motion with respect to the stellar wind plasma, we have an analogous situation to the magnetic planets in the Solar System, in which electrons would spiral along the magnetic field lines of the {planet}. In  contrast,  in Sept.~2009, the motion of HD179949b is sub-Alfvenic, indicating that star-planet magnetic coupling can take place. If that is the case,  electrons can also  spiral along the magnetic field lines connected to the {star}. If the criteria for generating ECM emission are met, in the first and second scenarios, the emission would originate from the planetary field, whereas in the second scenario, it could also originate from the stellar field.

\begin{figure}[t]
\includegraphics[width=1.2\textwidth]{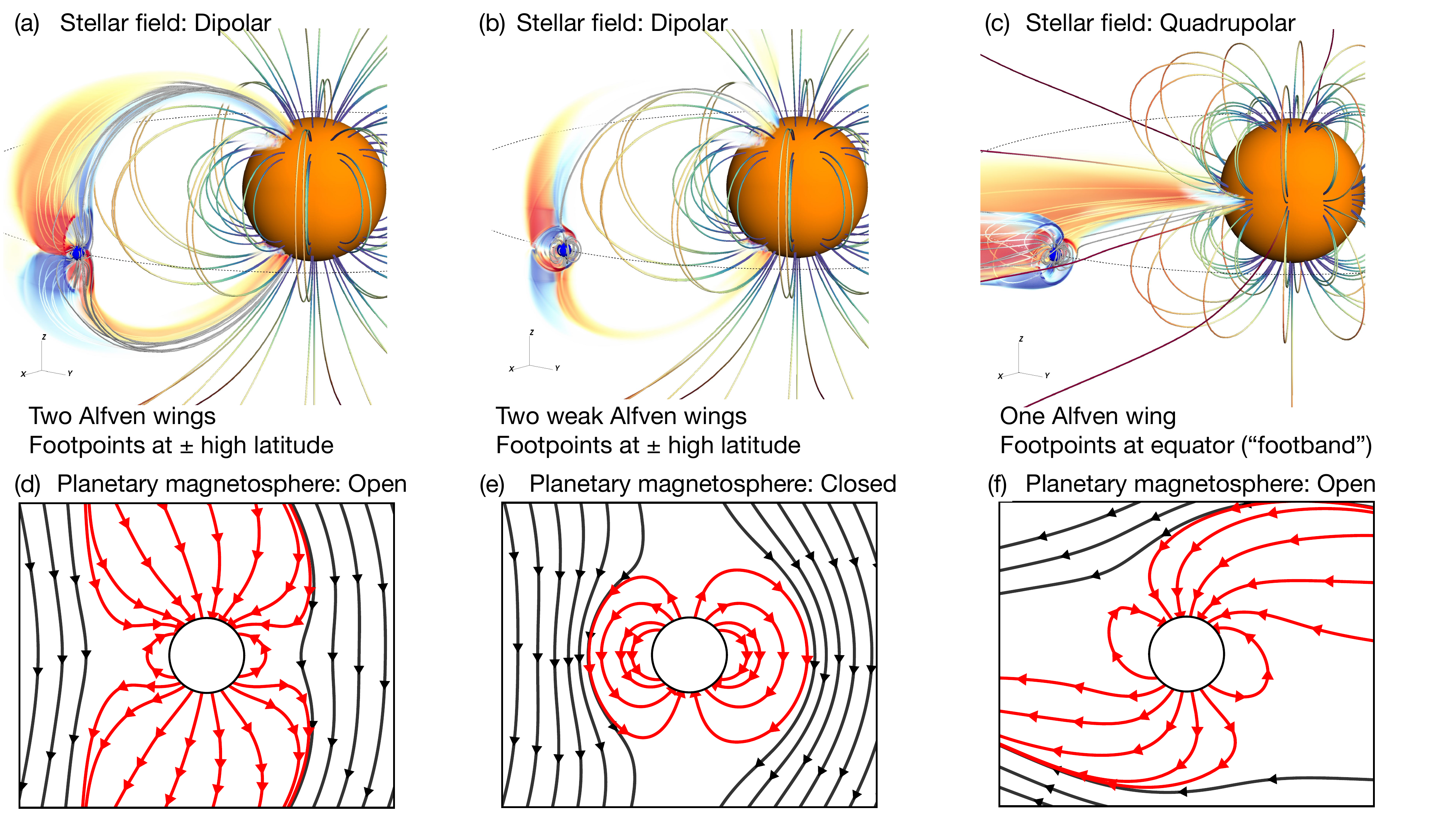}
\caption{Top row: The effects of magnetic field topologies for planet and star on the formation of Alfven wing currents, shown in red/blue colours. The planet and the star are represented by the blue and orange spheres. Bottom row: Zoom-in showing the planetary (red streamlines) and stellar (black streamlines) magnetic field lines corresponding to the simulations shown directly above in the top row. 
The footpoints of current-carrying field lines reach the stellar surface at different latitudes, depending on the magnetic field topologies. (a/d) Planetary and stellar surface dipolar magnetic moments are anti-parallel. (b/e) Planetary and stellar surface dipolar magnetic moments are parallel. (c/f) Surface planetary magnetic moment is as in panel a and a quadrupolar stellar field. Figure adapted  from \citet{2015ApJ...815..111S}, copyright AAS. Reproduced with permission. }\label{fig.strugarek} 
\end{figure}

\begin{marginnote}
\entry{Open magnetosphere}{refers to a magnetosphere in which a fraction of the magnetic field lines are opened due to star--planet magnetic field reconnection.}
\end{marginnote}

Magnetic coupling is most effectively studied with 3D models. This is because the Poynting energy flux carried by the waves from the planet to the star depends on the 3D topology of the stellar and the planetary magnetic fields \citep{2015ApJ...815..111S, 2023A&A...671A.133E}. Figure \ref{fig.strugarek} shows the results of three simulations of magnetic SPI from  \citet{2015ApJ...815..111S}. Panels a-c show  the Alfven wing currents that propagate parallel to the magnetic field. {Panels d-f show the topologies of planetary and stellar  magnetic fields corresponding to the simulations shown in Figure \ref{fig.strugarek}a-c, respectively. Although the three simulations assume that the planetary magnetic field is dipolar, the dipolar magnetic moment in panel e has a positive vertical orientation, whereas panels d and f have a negative vertical orientation. The magnetosphere of the planet can have a fraction of its magnetic field lines open when there is reconnection between the stellar and planetary field lines. For example, if the star and the planet have dipolar fields, and their respective surface magnetic moment are anti-parallel to each other, at the location where the field lines interact, they  have  opposite signs, thus leading to magnetic field reconnection and an open magnetosphere (Figure \ref{fig.strugarek}d). In this idealised scenario of purely dipolar fields, for a closed magnetosphere to occur, the surface dipolar magnetic moments of the planet and of the star must be parallel (Figure \ref{fig.strugarek}e).} 

Depending on the stellar and planetary field topologies, the Alfven wing currents (panels a-c) impinge the stellar surface at different locations and also with different extensions.  In the case where the planetary and stellar magnetic fields are dipolar (panels a/d and b/e), the footpoints of current-carrying field lines are located at high latitudes at the stellar surface. Note here also that the open planetary magnetosphere (panel a) generates two strong currents, whereas in the case of the closed planetary magnetosphere (panel b), the currents are much weaker. Poynting fluxes are significantly weaker when planetary magnetospheres are closed \citep{2023A&A...671A.133E}. Panels c/f illustrate another case where the  planetary magnetosphere is open. Here, the surface planetary dipolar magnetic moment has a negative vertical orientation (as in panel a), but the stellar magnetic field is quadrupolar. Because of this  topology, only one Alfven wing connecting to the star develops in the system. The wing currents now reach the stellar surface at the equator and, instead of being  spot-like distributed  as in panel a, it now resembles a band.

The model presented in Figure \ref{fig.strugarek}, from \citet{2015ApJ...815..111S}, includes the magnetised stellar wind, but neglects planetary outflows. In the next section, I will explore the models that describe flow-flow interactions, i.e., including both stellar and planetary outflows.

%%%%%%%%%%%%%%%%%%%%%%%%%%%%%%%%%%%%%%%%%%
\section{3D MHD MODELS OF FLOW-FLOW INTERACTIONS}
%%%%%%%%%%%%%%%%%%%%%%%%%%%%%%%%%%%%%%%%%%%%%%%%%%%%%%%%%%%%%%%%
\subsection{Morphological classification of magnetised SPIs}
A number of 3D simulations  investigated the morphology of dynamical interactions between the outflow of a close-in planet and the stellar wind. I highlight here the works of \citet{2015A&A...578A...6M} and  \citet{2021Univ....7..422Z}, who performed 3D simulations of the stellar wind interacting with escaping atmospheres of close-in planets. These works utilise global simulation setups and the planetary outflow is treated in a more simplified way (i.e., the energetics of photoevaporation are not included in these models, see Section \ref{sec.pl_outflow}). Both works consider the effects of stellar and planetary magnetic fields. Although  \citet{2015A&A...578A...6M} mostly dealt with super-Alfvenic interactions, \citet{2021Univ....7..422Z} investigated  interactions that are sub-Alfvenic and interactions that they call trans-Alfvenic, in which the stellar wind is sub-Alfvenic at the orbit of the planet, but because of  the extra velocity component due to orbital motion, in the reference frame of the planet, the stellar wind impinges the planet at mildly super-Alfvenic velocities.

\begin{figure}[t]
\includegraphics[width=0.91\textwidth]{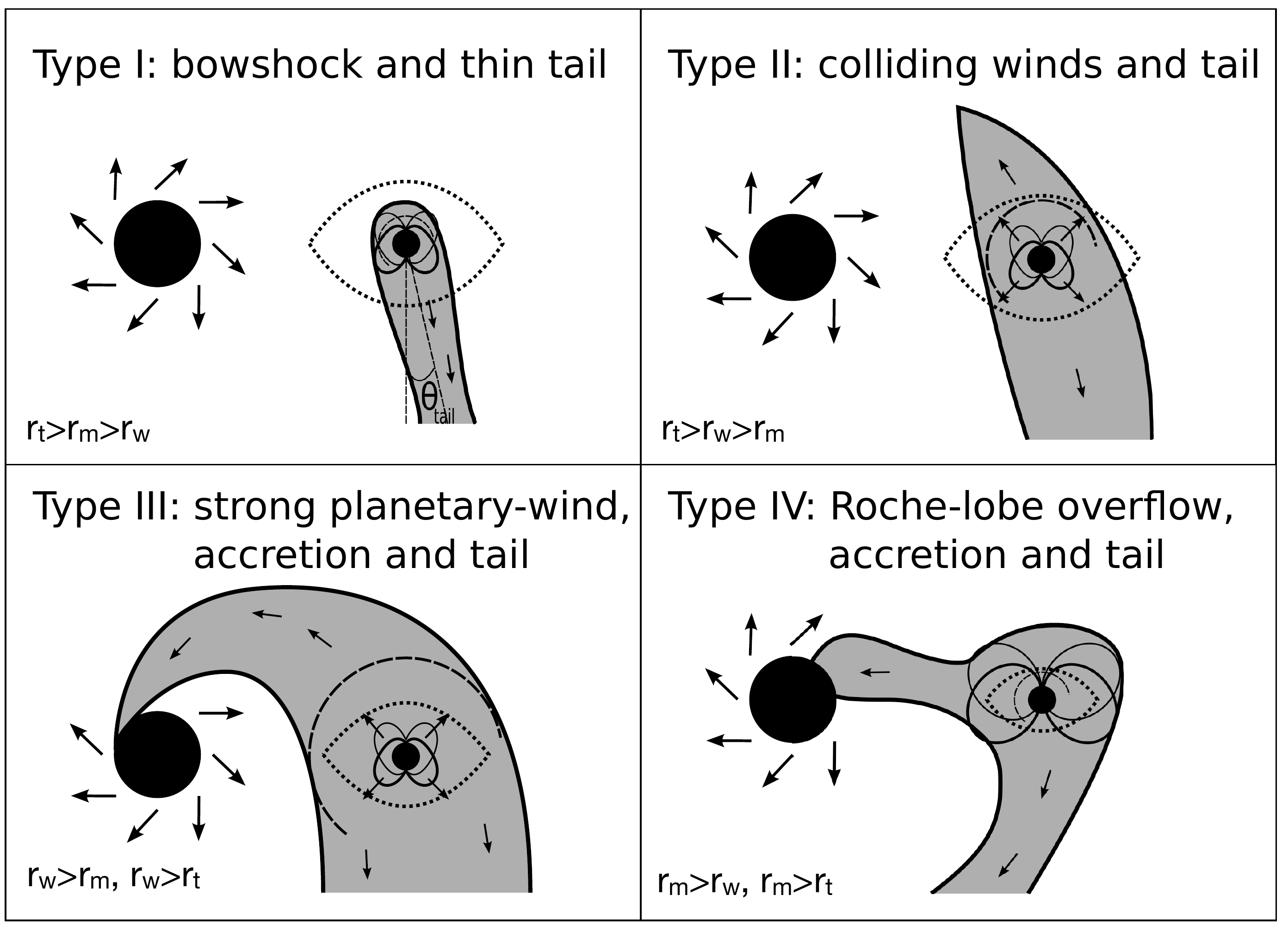}\\
\includegraphics[width=1.2\textwidth]{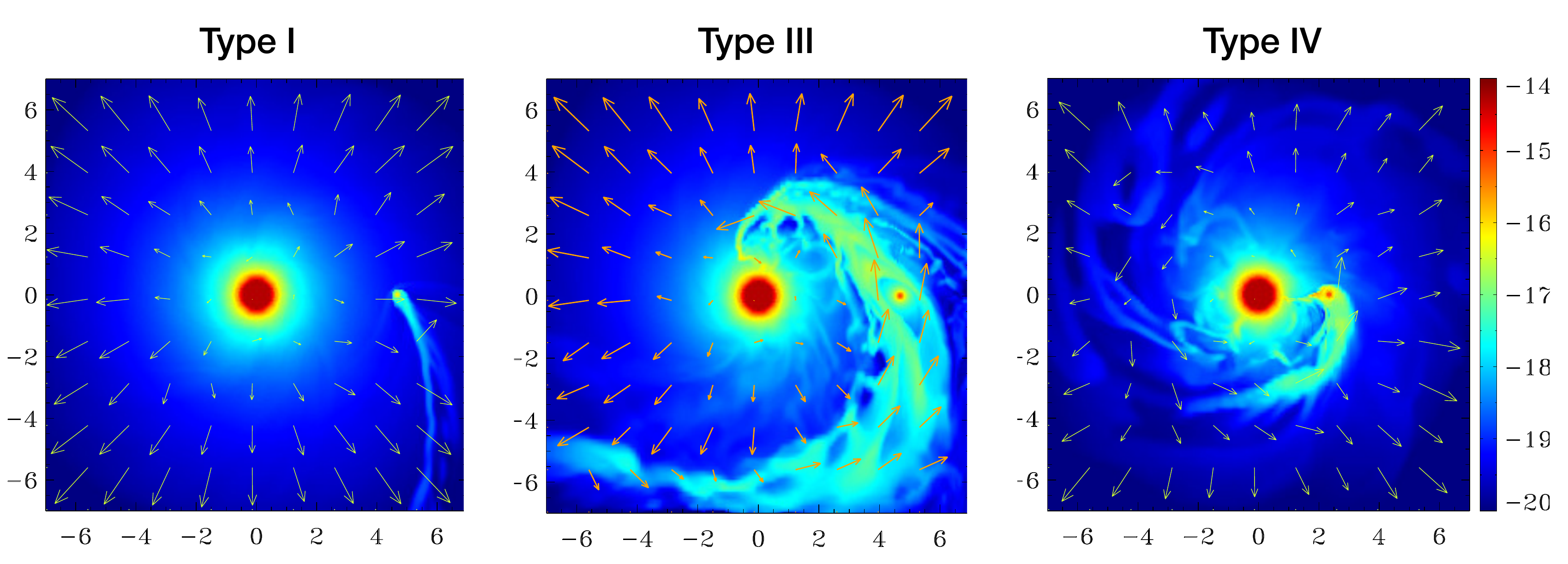}\\
\caption{General morphological classification of magnetised flow-flow interactions. The eye shaped dotted line represents the equipotential where the inwards gravitational force balances the outwards centrifugal force, with $r_t$ being the distance from the centre of the planet to the cusp closest to the star (corresponding to the Lagrangian L1 point). The dashed arcs represent $r_w$ and the dipolar field lines represent the extension of the magnetosphere $r_m$. (a) The large-scale structures based on the relative distances among $r_m$, $r_t$ and $r_w$. (b) Equatorial cuts of three different simulations where the morphological structures named Type I, III and IV have been identified. None of the simulations from \citet{2021MNRAS.500.3382C} included the Type II morphology, but this morphology is similar to the one shown in Figure \ref{fig.carolan21}a from the work of \citet{2021MNRAS.500.3382C}.  Figure adapted with permission from \citet{2015A&A...578A...6M}; copyright ESO. \label{fig.matsakos} }
\end{figure}

Figure \ref{fig.matsakos} summarises the general morphological classification of the interaction between a magnetised stellar wind and a magnetised close-in planet proposed by  \citet{2015A&A...578A...6M}. This diagram illustrates the large-scale structures that were recovered in their 3D simulations and their classification depends on three characteristic length-scales. The first one is the size of the magnetospheric stand-off distance of the planet $r_m$, which is located at the point where the ram pressure of the stellar wind balances the magnetic pressure of the planetary magnetosphere. The second  length-scale is the flow-flow stand-off distance $r_w$, which is the position where the stellar wind ram pressure and the planetary outflow ram pressure balance out. The third  length-scale is the tidal distance $r_t$, located where the effective gravity is zero, i.e., the inwards gravitational force balances the outwards centrifugal force. At the bottom left of each panel in Figure \ref{fig.matsakos}, I annotated the relative distances among $r_w$, $r_m$ and $r_t$ for each morphological type. Based on these relative distances, four types of morphologies were identified by \citet{2015A&A...578A...6M}:
\begin{marginnote} 
\entry{Magnetospheric stand-off distance $r_m$}{given by pressure balance between planetary magnetosphere and stellar wind.} 
\entry{Flow-flow stand-off distance $r_w$}{given by pressure balance from the two flows.}
\entry{Tidal distance  $r_t$}{where the inwards gravitational force is balanced by the outwards centrifugal force (i.e., effective gravity is zero).
}
 \end{marginnote}
 
\begin{enumerate}
\item[(a)] {Type I $\mathbf{(r_t>r_m>r_w)}$:} In this case, the  planetary outflow is weak and the stellar plasma is balanced by the planetary magnetic pressure. The planetary outflow does not expand beyond the Lagrangian point L1, and it is shaped in the form of a thin comet-like tail trailing the planet. As the stellar wind is super-Alfvenic at the interaction region, there is also formation of a bow shock ahead of the planet. An example of this type of interaction is shown in the bottom left panel of Figure \ref{fig.matsakos}. Here the star is shown at the origin of the grid, and the planet is at coordinate $[5,0]$ in simulation units. The arrows depict the stellar wind velocity and the colour indicates the logarithm of the total mass density. Note that both the orientation of the shock and the comet-like tail depend on the relative motion between the planet and the stellar wind \citep{2010ApJ...722L.168V}.

\item[(b)] {Type II $\mathbf{(r_t>r_w>r_m)}$:} In this case, the planetary outflow is sufficiently strong for it to blow open the magnetic field lines of the planet's magnetosphere. This outflow is balanced by the incoming stellar plasma, but the interaction takes place below the L1 point ($r_w<r_t$). Because of this, the planetary outflow stays in the vicinity of the orbit and it is not accreted by the star. In the parametric study of \citet{2015A&A...578A...6M}, none of their models reproduced this morphology type, but we can see an example of such morphology in Figure \ref{fig.carolan21}a extracted from the work of \citet{2021MNRAS.500.3382C}. In the numerical study of \citet{2024A&A...683A.226C},  this morphological type was found to be the most common  for the hot Jupiter HD189733b. Note that the colliding flows can also develop a shock and the comet-like tail is less collimated than that of morphology Type I. 

\item[(c)] {Type III $\mathbf{(r_w>r_m, \, r_w>r_t)}$:} In this case, the planetary outflow is also strong and, in addition, it overflows the Lagrangian L1 point. When this happens, the material is no longer gravitationally bound to the planet, and instead is bound to the stellar gravity. Subsequent interaction with the stellar wind removes angular momentum of the material, which spirals in towards the star. An interesting feature of this morphology type is that  accretion takes place ahead of the sub-planetary point. The middle bottom panel in Figure \ref{fig.matsakos} shows material being accreted at a $\sim 90$-degree phase lag -- the planet is again located at coordinate $[5,0]$ in simulation units and the accretion point takes place around coordinate $[0,1]$.

\item[(d)] {Type IV $\mathbf{(r_m>r_w, \,r_m>r_t)}$:} The final morphological type identified in the work of \citet{2015A&A...578A...6M} has the magnetosphere of the planet extending beyond L1 ($r_m>r_t$), such that material within the magnetosphere is attracted by the stellar gravity. Here, accretion of planetary material towards the star happens in a similar way as to a classical Roche lobe overflow (albeit with dynamics modified by magnetism). Contrary to the previous Type III morphological type, because of a relatively weak planetary outflow, atmospheric material does not spiral in towards the star and instead falls at the sub-planetary point (there is no phase lag). The right bottom panel in Figure \ref{fig.matsakos} shows material being accreted at around coordinate $[1,0]$, with the planet  located at coordinate $[5,0]$ in simulation units. 
\end{enumerate}

In Type III and Type IV morphologies, some material is accreted by the star and could therefore potentially generate hot spots in the host-star's chromosphere. In Section \ref{sec.observations}, I presented some earlier mechanisms proposed to explain anomalous stellar activity. The accreting material scenario seen in Type III and Type IV morphologies are alternative mechanisms that could also  explain anomalous stellar activity (see Section \ref{sec.accretion}  for more quantitative arguments). Additionally, we notice also in some morphological types (e.g., Type IV) that there is left-over planetary material surrounding the system, which can potentially form a torus of material surrounding the host star. If sufficiently optically thick, this material could absorb stellar light. This situation has been proposed to explain the absorption of the cores of the Mg II h\&k lines in the WASP-12 system \citep{2013ApJ...766L..20F}, and a similar morphology has been seen in the 3D HD models of \citet{2018MNRAS.478.2592D} and \citet{2019MNRAS.483.2600D}, using  global simulation setups.

%%%%%%%%%%%%%%%%%%%%%%%%%%%%%%%%%%%%%%%%%%
\subsection{Radiative MHD models of flow-flow interactions}\label{sec.pl_outflow}
The types of studies presented in the previous section are important for understanding the global properties of flow-flow interactions, and the effects that these different morphological structures can have on observational diagnostics, such as generation of chromospheric/coronal hot spots, formation of a circumstellar torus, formation of a bow shock, etc. However, these models are more limited in understanding the physical properties of atmospheric evaporation, as they lack a more robust treatment of the escaping planetary outflow. 

There are many escaping routes for a planet to lose its atmosphere \citep{2020JGRA..12527639G}. In the context of close-in exoplanets, hydrodynamic escape through photo-evaporation is likely one of the dominant escape processes. This happens because, due to geometrical effects, close-in planets receive large levels of irradiation from their stars. Irradiation more energetic than 13.6~eV is capable of ionising neutral hydrogen, with the excess energy being used to generate heating, which causes  atmospheres to inflate and make them more likely to outflow hydrodynamically. As discussed in Section \ref{sec.transmission}, transmission spectroscopy in the  Ly-$\alpha$ line performed with Hubble observations of transiting systems have revealed the presence of evaporating upper atmospheres of close-in giant planets. An important feature observed in Ly-$\alpha$ transits is that the blue wing of the line extends to blue-shifted velocities of a few hundreds of km/s. To explain these high velocities, neutral particles must be accelerated to such velocities. However, typical speeds of hydrodynamically escaping planetary outflows are only a few tens of km/s. It is believed that the processes responsible for generating high-velocity neutrals have their origins on the interaction with the host star's wind and possibly also its radiation (see green box in page \pageref{greenbox_wings}), and thus requires multi-dimensional models of flow-flow interactions for their interpretation.

\begin{textbox}[h]%
\section{WHAT CAUSES THE HIGH VELOCITIES OBSERVED AT THE WINGS OF LY-$\alpha$ LINES} \label{greenbox_wings}%
There are three proposed scenarios to explain the high velocities observed at the blue wings of Ly-$\alpha$ lines. 
\begin{enumerate}
\item The first explanation is that  the stellar outflow drags along planetary material as it passes through the planet, accelerating substantial amounts of escaping neutrals towards the observer, required to produce enough optical depth in the blue wing of the Ly-$\alpha$ line \citep{2019ApJ...873...89M, 2021MNRAS.501.4383V}. This scenario often requires a strong wind or even the passing of a CME  \citep{2022MNRAS.509.5858H}. 

\item The second suggestion also invokes the presence of the stellar wind, but here  charge-exchange processes between the neutral planetary outflow and the ionised stellar wind lead to neutralisation of fast stellar wind protons, thus contributing to high-velocity blue-shifted absorption. The effectiveness of this process is debated in the literature, with substantial \citep{2013MNRAS.428.2565T}, mild  \citep{2019MNRAS.487.5788E} and no effects \citep{2022MNRAS.517.1724D} caused by charge-exchange processes found even for the same system.

\item A third suggestion is that the neutral atmosphere of the planet is accelerated by Ly-$\alpha$ radiation pressure \citep{2013A&A...557A.124B}. This suggestion is  less supported from the perspective of fluid models \citep{2018MNRAS.475..605C, 2020MNRAS.493.1292D}. 
\end{enumerate}
In reality, it is very likely that two or more of these mechanisms operate simultaneously in the system, all contributing to the enhancement of blue-shifted absorption in the Ly-$\alpha$ line \citep{2019ApJ...885...67K, 2019MNRAS.487.5788E}. 
\end{textbox}

For clarity, I divide here the models investigating flow-flow interactions depending on their treatment of the acceleration of planetary outflows. In general, models that cover  larger spatial scales and therefore use the global simulation setups, tend to use simplified treatments for the planetary outflow, neglecting for example more complex process in the acceleration of the escaping atmosphere. Examples of such setups are found in the studies from \citet{2014MNRAS.438.1654V, 2018MNRAS.479.3115V, 2015A&A...578A...6M,  2017MNRAS.466.2458C, 2021Univ....7..422Z}, including the very first 3D star-planet flow-flow model from \citet{2007ApJ...671L..57S}.  One notable exception is the work of \citet{2019ApJ...885...67K}, which in spite of using a global simulation setup, also implements a more physically realistic acceleration of planetary outflows. In general,  3D models that provide better treatment of the acceleration of planetary outflows tend to use local simulation setups (although many local simulations can also use simplified treatments of outflow driving). The initial 3D studies to self-consistently compute  the launching of the planetary outflow through photo-evaporation were developed by \citet{2018MNRAS.481.5315S, 2019MNRAS.483.1481D, 2019ApJ...873...89M}. These types of models, which include photo-ionisation driving, are also known as radiative (M)HD models, as the radiative transfer of the incoming stellar  high-energy  irradiation is self-consistently  built in the simulation setup. It is only very recently that the first radiative MHD models, i.e., with self-consistent outflow acceleration including the effects of planetary magnetic fields \citep[][global simulation setup]{2021MNRAS.507.3626K} and also of magnetised stellar winds \citep[][local simulation setup]{2021MNRAS.508.6001C} were published.

%%%%%%%%%%%%%%%%%%%%%%%%%%%%%%%%%%%%%%%%%%%%%%%%%%%%%%%%%%%%%%%%%
\section{SCIENCE HIGHLIGHTS OF FLOW-FLOW INTERACTION MODELS}
Models used to investigate flow-flow interactions (HD/MHD or radiative-HD/-MHD) can vary substantially in the numerical codes they use, the details of the included processes, etc. Ultimately, the choice of simulation setup as well as the complexity of the models rely on the scientific question being addressed. In this section, I provide some scientific highlights and lessons learned from various 3D studies.

\subsection{The influence of stellar wind strength on the observability of atmospheric escape in Ly-$\alpha$} 
In Section \ref{sec.subcritical}, I showed that planetary mass-loss rate is reduced if the interaction with the stellar wind takes place within the sub-critical region of the planetary outflow (Figures \ref{fig.pl_conditions} and \ref{fig.carolan21}). If the interaction happens when the planetary outflow already has super-critical velocities, the stellar wind does not affect escape rates. However, even when this happens, there can still be a strong effect of the wind on observational signatures from spectroscopic transits \citep[e.g.,][]{2014MNRAS.438.1654V}. An obvious effect of the stellar wind is on the confinement of an expanding atmosphere within a certain volume. The larger the wind pressures, the smaller  the volume that such atmosphere can occupy (this is seen as one goes from left to right panels in Figure \ref{fig.carolan21}). The limited volume occupied by the atmosphere changes the optical depth of the atmosphere and, as the planet transits and stellar photons are transmitted through the atmosphere of the planet, this affects the transit signature in the Ly-$\alpha$ line.

\begin{figure}[t]
	\centering
	\includegraphics[width=1.2\textwidth]{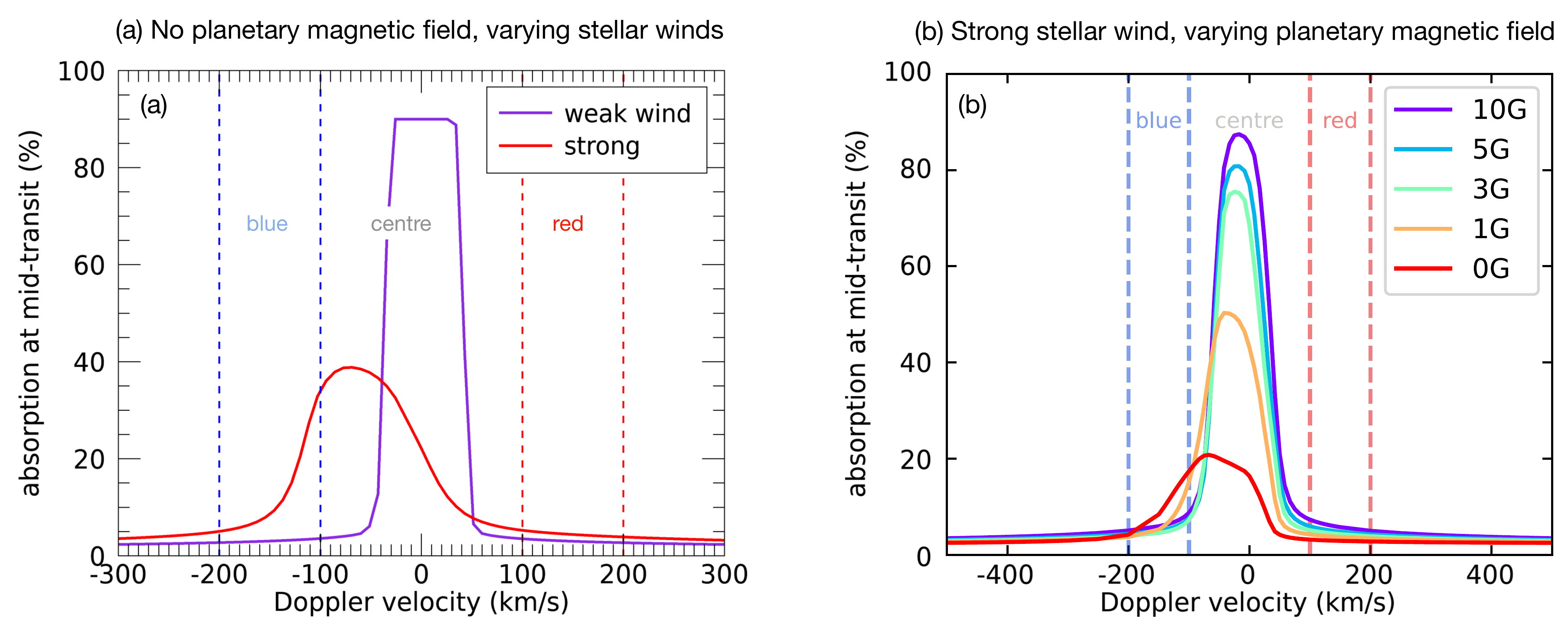}	
	\caption{Ly-$\alpha$ line synthetically obtained by transiting a close-in gas giant with an evaporating atmosphere under different stellar wind conditions and magnetisations: (a) Considering no planetary magnetic field, with varying stellar wind strengths.  (b) Considering a moderate stellar wind, with varying planetary magnetic field strengths.  Note here that the models shown in these panels were computed for planets with different irradiation levels, which is why their absorption levels are not identical. Panel $b$ adapted from \citet{2021MNRAS.508.6001C}; copyright 2021 The Author(s).}\label{fig.synthetic_lines_b}
\end{figure}

The influence of stellar wind strength on the observability of atmospheric escape in Ly-$\alpha$ has been the topic of several 3D simulation studies, such as: \citet{2014MNRAS.438.1654V, 2019ApJ...873...89M, 2020A&A...639A.109S, 2020MNRAS.498L..53C, 2021MNRAS.500.3382C, 2021MNRAS.507.3626K}. These studies demonstrated that, in the case of a dynamically weak stellar wind,  most of the absorption of the planetary atmosphere takes place closer to the planet. This is the region where material is still being accelerated, and consequently still have small velocities, thus generating absorption close to the centre of the Ly-$\alpha$ line. One drawback of  Ly-$\alpha$ observations is that the interstellar medium and geocoronal emission obscure the centre of the line, so the line centre cannot be used to probe atmospheric evaporation. However, in the presence of a dynamically stronger stellar wind, more planetary material is pushed towards the comet-like tail. Because this material can reach high blue-shifted velocities, the Ly-$\alpha$ line becomes more skewed to negative Doppler velocities. These two considerations are illustrated in Figure \ref{fig.synthetic_lines_b}a, where I plot the Ly-$\alpha$ line synthetically obtained by transiting a close-in gas giant with an evaporating atmosphere, considering two wind conditions. The contaminated region is labeled as `centre' in Figure \ref{fig.synthetic_lines_b}a, where I also indicate the blue and red wings with their respective labels. 

The wings of the  Ly-$\alpha$  line do not suffer from this contamination and are therefore  important diagnostics of atmospheric escape and also of the stellar wind environment \citep{2017MNRAS.470.4026V}.  Because of the great amount of hydrogen atoms escaping, even at great distances from the planet where material is substantially ionised, there is still a significant amount of neutral hydrogen to absorb the wings of the  Ly-$\alpha$  line (see Section \ref{sec.pl_outflow} and Figure  \ref{fig.synthetic_lines_b}a). Because of the asymmetric distribution of material around the planet, the Ly-$\alpha$  line is not symmetric about line centre, showing different amounts of absorption in the blue and red wings of the  Ly-$\alpha$  line. Furthermore,  there is also temporal asymmetry in the lightcurve, i.e., differences in the pre-ingress phase and in the after-egress phase. This is because the distribution of material ahead of the planetary orbit (which starts transiting before the transit of the geometric disc of the planet) and material trailing the planet (which transits after the transit of the geometric disc of the planet) is not symmetric, and signals can still occur far outside of the geometric transit \citep[e.g.,][]{2019ApJ...873...89M}. 

\subsection{Evaporated material from a close-in planet can be accreted onto the host star generating a localised hot spot} \label{sec.accretion}
\begin{figure}[t]
\includegraphics[width=0.99\textwidth]{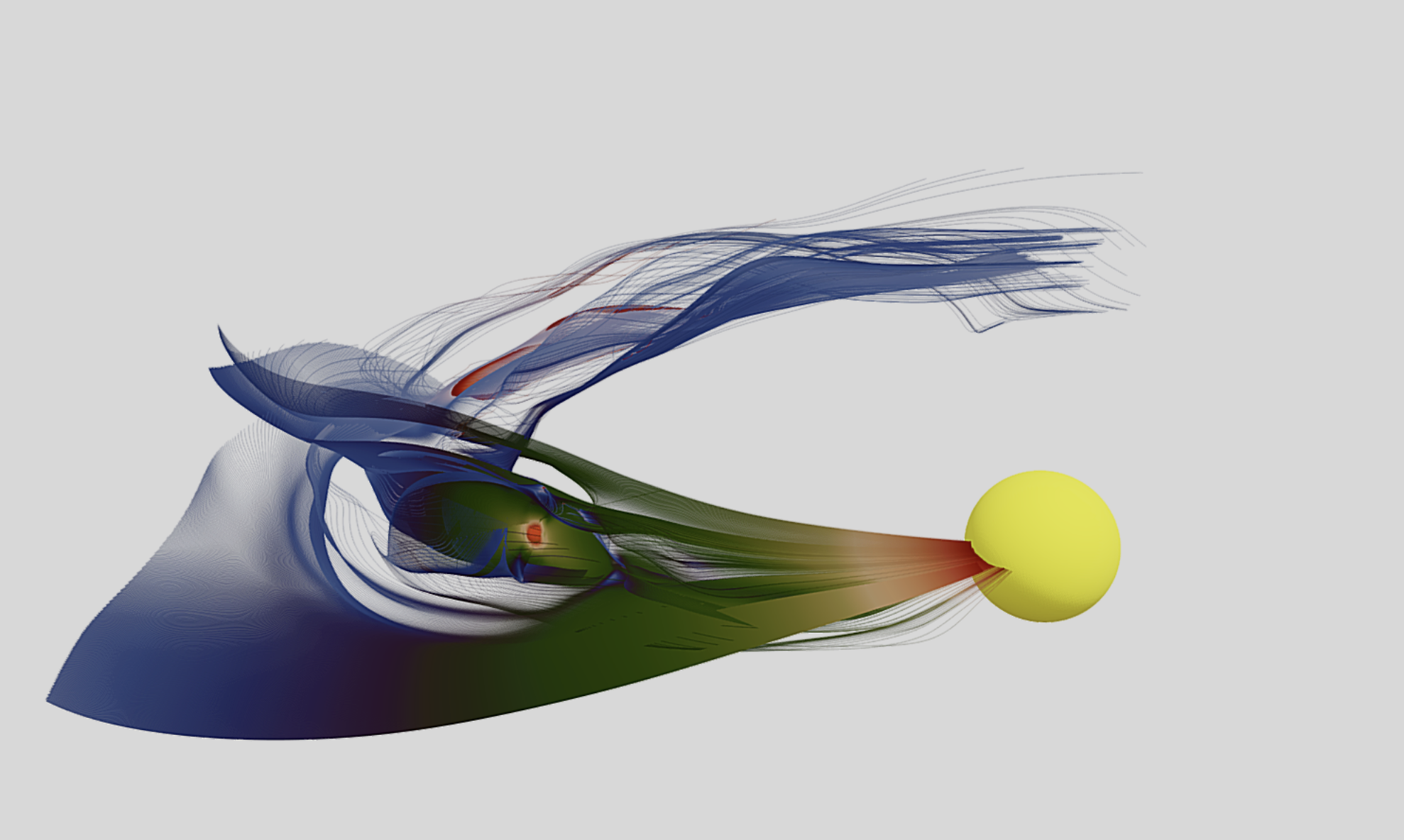}
\caption{Rendering of the planetary and stellar magnetic field lines. The planet and the star are represented by the red and yellow spheres, respectively. The position at the stellar surface where material is accreted depends on the topology of the stellar magnetic field. Figure adapted from \citet{2024A&A...683A.226C} (CC BY 4.0).  \label{fig.accretion} }
\end{figure}

Anomalous chromospheric/coronal activity detected in some exoplanetary systems has been explained with a number of SPI mechanisms that could excite the formation of hot spots on the stellar host. Some earlier suggested mechanisms included the formation of tidal bulges, and star-planet magnetic reconnection events (see Section \ref{sec.observations}). Another mechanism that could explain the formation of hot spots is through planetary evaporated material being accreted onto the central star, which was qualitatively presented in \citet{2015A&A...578A...6M}. To quantify this idea, a few authors performed 3D MHD models of flow-flow interactions, confirming that in some cases planetary material is indeed capable of creating a hot spot at the stellar surface  \citep{2019MNRAS.483.2600D, 2024A&A...683A.226C}. In the work of \citet{2019MNRAS.483.2600D}, who assumed a dipolar stellar magnetic field topology, they found that the planetary material spiralling onto the star climbs the closed stellar magnetic field lines, until it intersects the stellar surface.  This accretion phenomenon bears resemblance to magnetospheric accretion from disc material in classical T Tauri stars \citep{2005ApJ...634.1214L}. 
Similar to young stars' magnetospheric accretion, the location where planetary material impacts on the host star depends on the geometry of the stellar magnetic field. In Figure \ref{fig.accretion}, we see a 3D rendering from \citep{2024A&A...683A.226C} -- here, the stellar magnetic field line is fully open, giving rise to a Parker spiral. In this case, material that accreted onto the star would arrive at equatorial latitudes (the dipolar topology seen in \citealt{2019MNRAS.483.2600D} leads to hot spots  located at higher latitudes in the stellar surface). Although accretion from planetary material can take place,   \citet{2019MNRAS.483.2600D} and \citet{ 2024A&A...683A.226C} showed that the emission generated by the hot spot is likely too small to be disentangled from stellar activity. 

Although some of the mechanisms suggested to explain anomalous stellar activity require magnetic coupling between the star and the planet (Section \ref{sec.connectivity}), the simulations  in \citet{2019MNRAS.483.2600D} are super-Alfvenic (i.e., with no direct magnetic coupling between the stellar and planetary magnetic fields). In this case, it is the effect of  material falling at the star that leads to the formation of the hot spot. With the use of passive scalars, \citet{2019MNRAS.483.2600D} were able to trace the material that is accreted through the hot spot. They showed, nevertheless, that only about 5\% of the total amount of mass lost by the planet ends up at the surface of the star;  most of the material being accreted was in fact stellar wind material which, after interaction with the planetary outflow, has fallen back to the star. \citet{2019MNRAS.483.2600D} and \citet{2024A&A...683A.226C} showed, furthermore, that planetary magnetism cannot be ignored, as even modest field strengths can affect  substantially the wind structure and the accretion stream. \citet{2024A&A...683A.226C}  showed that instabilities generated at the contact surface between planetary and stellar outflows are important to generate the accreting inflows. Next, I will present another potential effect that planetary magnetic fields can have on SPIs, namely how it can affect atmospheric evaporation (see the sidebar titled Planetary Magnetism: Does It Reduce or Enhance Atmospheric Escape?. 

\begin{textbox}[h]%
\section{PLANETARY MAGNETISM: DOES IT REDUCE OR ENHANCE ATMOSPHERIC ESCAPE?}
Planets that are weakly or non-magnetised lack a substantially large protective bubble around them, which implies that their atmospheres become exposed to the impact of stellar outflows. Due to sputtering processes,  stellar winds may then remove their atmospheres. Mars is an example of such a planet in the Solar System \citep{2007SSRv..129..207K} and this process might take place in unmagnetised exoplanets as well \citep{2018MNRAS.481.5296V}. For this reason, planetary magnetic fields are believed to offer favourable conditions for planetary habitability \citep[e.g.][]{2016pmf..rept.....L, 2018haex.bookE..57M}. However, some authors have argued that magnetic fields do not prevent atmospheric escape, with Mars/Earth being counter examples of unmagnetised/magnetised planets with similar outflow rates. This is currently being debated, with views in favour \citep{2014PEPI..233...68T, 2018MNRAS.481.5146B} and against \citep{2018A&A...614L...3G}  protective effects, and models showing that atmospheric protection depends on magnetisation \citep{2019MNRAS.488.2108E, 2022JGRA..12730427S}.

Most of this discussion relating to the protective effects of planetary magnetic fields has revolved around potentially habitable terrestrial planets, whose atmospheres undergo escape at much lower rates than what we observe in close-in gas giants. It is  not clear whether observations of mass loss in terrestrial Solar System planets can be extended to close-in giant exoplanets, whose ambient and physical conditions differ so greatly.  Additionally, the escape processes in these two families of planets are different. Whereas in the terrestrial planets non-thermal escape dominates (e.g., polar wind, cusp escape, ion pickup, \citealt{2018A&A...614L...3G}), in close-in gas giants, the  dominant escape process  is bulk hydrodynamic escape, which might have been important for the young Earth, but is negligible in all Solar-System planets today.
\end{textbox}

%%%%%%%%%%%%%%%%%%%%%%%%%%%%%%%%%%%%%%%%%%%%%%%
\subsection{The effects of planetary magnetic fields in atmospheric escape} \label{sec.magnetic_escape}
Even though there are many models investigating the effects of planetary magnetic fields in SPIs occurring in close-in gas giants, radiative MHD models that self-consistently include atmospheric escape driving through photo-evaporation and the presence of the planetary magnetic fields are still at their infancy. \citet{2021MNRAS.508.6001C} and \citet{2021MNRAS.507.3626K} developed the first 3D radiative MHD models that included both magnetisation and self-consistent planetary outflow driving. However, when investigating the effects of planetary magnetic fields in atmospheric escape, these models reached different conclusions.  

In the 3D parametric study of planetary outflows from a magnetised hot Jupiter, \citet{2021MNRAS.508.6001C} found that the escape rate weakly increases with field strength: varying the polar planetary field strengths from 0 to 10~G, these authors found that mass-loss rates increase by a factor of $2$.  In contrast, \citet{2021MNRAS.507.3626K} increased the equatorial planetary magnetic field strength from 0 to 1~G and  observed a decrease in escape rate of a factor of $2$. What causes their conflicting conclusions? One possibility is that the models of \citet{2021MNRAS.507.3626K} do not include the stellar wind magnetisation, which can open up some planetary magnetic field lines, thus allowing for planetary material to more easily escape, as seen in the work of \citet{2021MNRAS.508.6001C}. There are also other subtle but very important differences in the high-energy fluxes considered in both studies, as well as stellar wind mass-loss rates. Another possibility is that these simulations differ in the fluid treatment: Although \citet{2021MNRAS.508.6001C} use a multi-species but single-fluid approach, \citet{2021MNRAS.507.3626K} consider a multi-fluid model. These approaches are expected to produce similar results in the limit of high densities, as is potentially the case of a dense planetary outflow (in particular, in the lower part of the escaping atmosphere). In high-density regimes,  collisions between neutral and ion species are frequent and the whole material tends to move along the magnetic field lines. However, when collisions between the neutral and ionised species become less frequent, ions move along the magnetic field, but neutrals can flow across the field, being affected by the magnetic field only via collisions with the ions. To account for the effect of neutrals' diffusion across the magnetic field that naturally arises in multi-fluid treatments, single-fluid models usually incorporate ambipolar diffusion terms in the magnetic induction equation. At the time of writing, we still do not know where the differences stem from and further comparison between these models are necessary. It is interesting to note as well that the 3D models of \citet{2021MNRAS.507.3626K} showed a weaker effect of planetary magnetisation on atmospheric escape rates than that obtained from their earlier 2D models  \citep{2015ApJ...813...50K}. These 2D models showed a decrease of one order of magnitude in escape rates of magnetised planets, as compared to unmagnetised ones, versus a factor of 2 computed using  3D simulations \citep{2021MNRAS.507.3626K}, indicating that the effects of magnetism on planetary evaporation might be more effectively studied in 3D.

\begin{figure}[t]
	\includegraphics[width=.9\textwidth]{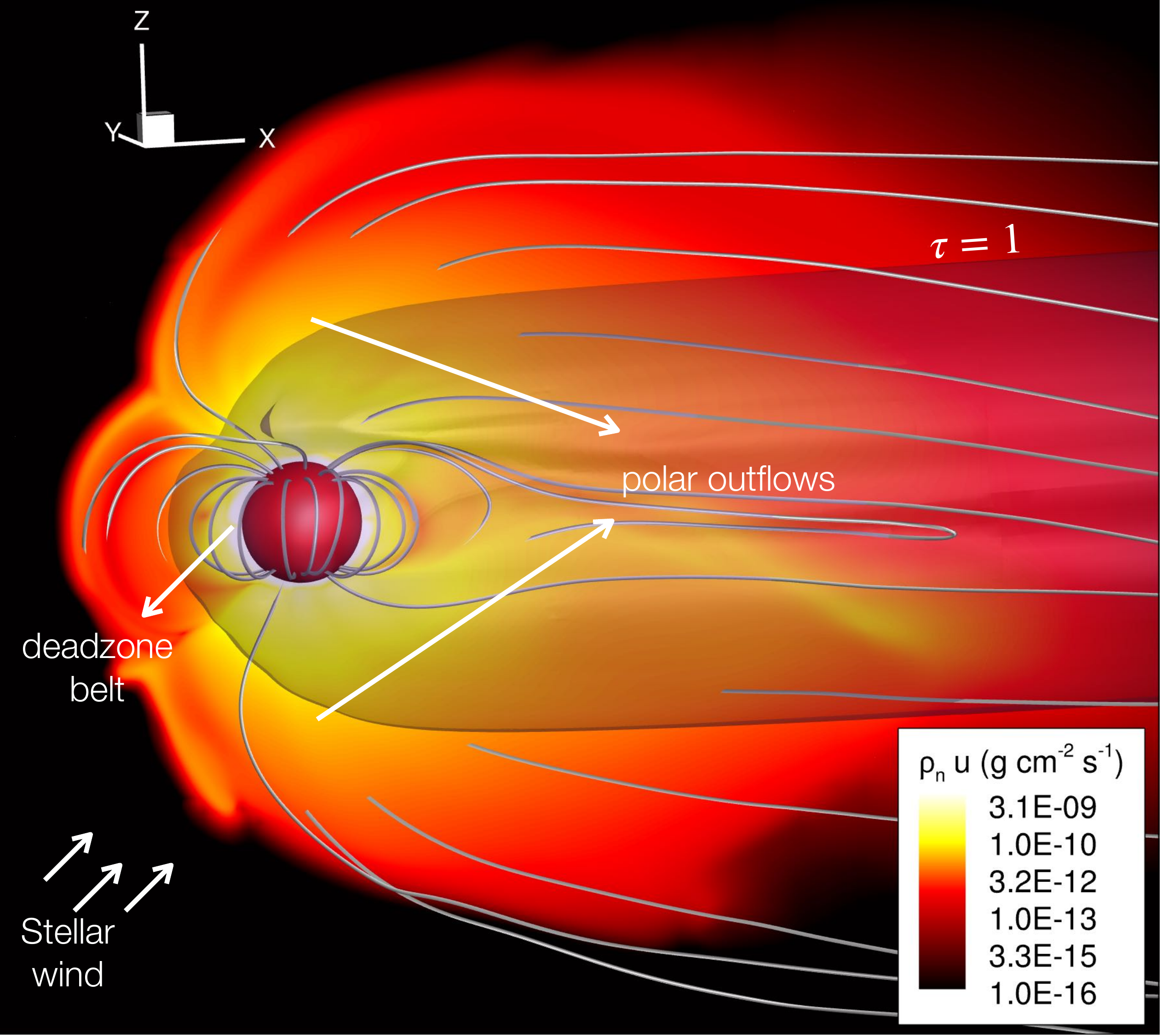}
	\caption{The planetary magnetic field alters the dynamics of atmospheric escape. Here, we see the formation of two tails emerging from each polar region, as traced by the atmospheric mass flux (colour). Weakly or non-magnetised planets generate instead one comet-like tail structure, trailing the orbital motion (cp.~Figure \ref{fig.local_global}). The grey surface shows where the optical depth for photoionisation of neutral hydrogen is $\tau=1$. This simulation assumes a planetary magnetic field of 10~G, and the super-Alfvenic stellar wind is injected from the $-x$ axis, similar to the model presented in  \citet{2021MNRAS.508.6001C}. 
	}\label{fig.mag_pl}
\end{figure}

The planetary magnetic field also changes the dynamics and the distribution of atmospheric material around the planet. In the case of un-magnetised planet, models predict that, as the atmosphere of the planet is lost and influenced by tidal and Coriolis forces and interact with the stellar wind, a comet-like tail of trailing atmosphere is formed similar to the models depicted in Figure \ref{fig.local_global}. However, as seen in Figure \ref{fig.mag_pl}, magnetised 3D models show the formation of dead-zones that do not participate in the mass-loss process and, in the case of super-Alfvenic interactions, two tails (one from each magnetic pole) trailing the planetary motion \citep{2021MNRAS.508.6001C}.\footnote{In the case of sub-Alfvenic interaction, the dynamics of escape change and escape from one magnetic pole is substantially less strong \citep{2024MNRAS.534.3622P}.} This is because magnetic fields, when sufficiently strong, funnel escape along the magnetic poles in the form of polar outflows. As a result, this  changes the distribution of material around the planet, which  then affects the resulting transmission spectroscopic transit.  

It has been demonstrated that  the magnetic field  substantially affects the amount of absorption observed during transits.   The main effect of a strong planetary magnetisation on the Ly-$\alpha$ line is to confine low-velocity material in the dead-zone, therefore enhancing the absorption at line centre. The effects of varying the planetary magnetisation on the Ly-$\alpha$ transit absorption can be seen in Figure \ref{fig.synthetic_lines_b}b, where we see that the stronger the planetary magnetisation is, the higher  the absorption at line centre. See also the sidebar titled Degeneracy Between Stellar Wind and Planetary Magnetisation. 

%%%%%%%%%%%%%%%%%%%%%%%%%%%%%%%%%%%%%%%%%%%%%%%%%%
\begin{textbox}[!b]%
\section{DEGENERACY BETWEEN STELLAR WIND AND PLANETARY MAGNETISATION}\label{greenbox_degeneracy}%
The effects of a strong planetary magnetic field and those of a weak wind on the Ly-$\alpha$ line are qualitatively similar, as shown in Figure \ref{fig.synthetic_lines_b}. In the case of unmagnetised planets (Figure \ref{fig.synthetic_lines_b}a),  a strong stellar wind can confine the escaping atmosphere to a smaller volume, reducing its overall absorption and shifting this absorption to blue-shifted velocities, due to the presence of a comet-like tail. In the case of a magnetised planet (Figure \ref{fig.synthetic_lines_b}b), the magnetic field  confines the atmosphere of the planet in the dead zone, forming a belt of material   around the planet -- the stronger the planetary magnetic field, the larger  the amount of material with low Doppler velocities, thus increasing line absorption at line centre. When comparing the panels in Fig.\,\ref{fig.synthetic_lines_b}, we see that there is a degeneracy between the effects of a strong (10-G) planetary magnetic field in panel b and those of a weak wind in panel a on the Ly-$\alpha$ line, as these effects produce qualitatively similar line profiles. One possible way to break this degeneracy is to have additional  studies that can independently either constrain   the stellar wind properties or characterise the planetary magnetic field. Alternatively, one could also use multiple spectral lines that diagnose escape at different altitudes. For example, the Ly-$\alpha$ line wings probe the region of interaction between the planetary upper atmosphere and the stellar wind, whereas heavier species are more likely to probe regions that are less affected by the stellar wind interaction and more affected by the confinement due to the planet's magnetic field.  Another possibility to break this degeneracy is to include in the analysis also the red-wing absorptions. Using 3D MHD models, \citet{2024MNRAS.534.3622P} showed that similar absorption in the blue wing could  be explained by either a highly magnetised planet under a weak stellar wind or a lowly magnetised planet under stronger stellar wind conditions. However, in the case of an atmosphere being blown by a strong stellar wind, the blue wing absorption can be a few times higher than the red wing absorption for any planetary magnetisation, whereas in the case of a weak stellar wind, the ratio between blue wing and red wing absorptions remain around unity.  
\end{textbox}

%%%%%%%%%%%%%%%%%%%%%%%%%%%%%%%%%%%%%%%%%%%%%%%%%%%%%%%%%%%%%%%%%
\section{FUTURE PROSPECTS AND MODEL REQUIREMENTS}\label{sec.future}

%%%%%%%%%%%%%%%%%%%%%%%%%%%%%%%%%%%%%%%%%%%
\subsection{Timescales of SPIs} \label{sec.timescales}
\begin{figure}[t]
\includegraphics[width=.99\textwidth]{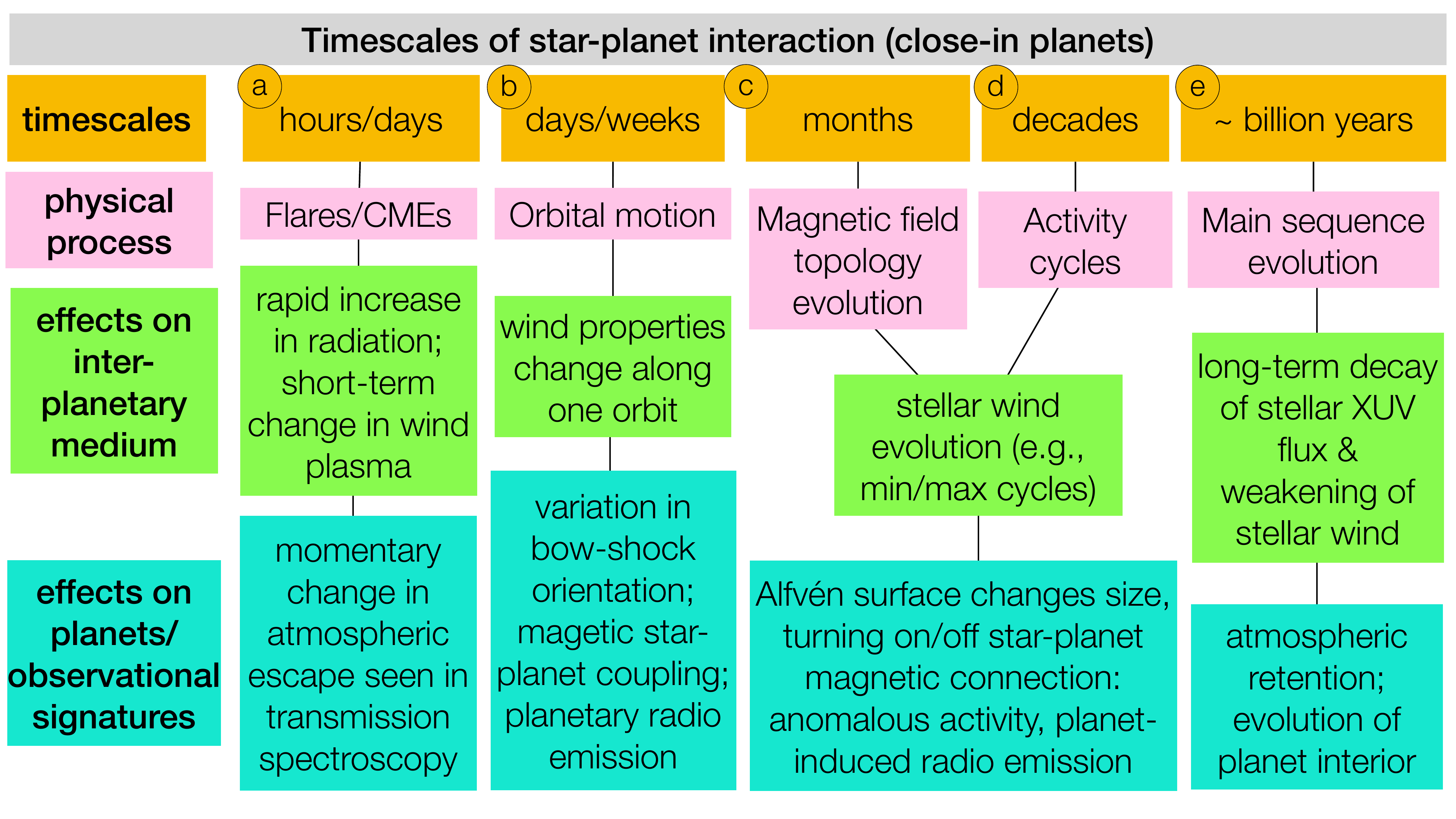}
\caption{Summary of timescales associated to different SPIs. Temporal evolution of SPI can be related to the evolution of stellar properties  (flares, CMEs, cycles) and planetary properties (evolution of planetary magnetisation, migration). The different timescales labeled a to e are discussed in the text. \label{fig.timescales}}
\end{figure}

Deciding on whether one needs a time-dependent model to simulate SPIs and the time-resolution of such models depends on the timescales one wants to investigate. Temporal changes in stellar magnetism lead to temporal changes in stellar particle ejecta and radiation and, consequently, in changes in the SPI. This can happen at various timescales and Figure \ref{fig.timescales} presents a summary of timescales associated to SPIs.

\begin{enumerate}
\item[(a)]  Over short timescales (on the order of hours/days) the rapid increase in energy input caused by flares can enhance, for example, atmospheric escape in exoplanets \citep[e.g.,][]{2020MNRAS.496.4017H} and potentially increase radio aurorae (see point b below). Additionally, if flares are accompanied by stellar CMEs, the associated particle flux of CMEs can affect the escape process momentarily \citep{2017ApJ...846...31C, 2020A&A...638A..49O, 2022MNRAS.509.5858H}. In fact, this short time variation might have been captured in spectroscopic transits of the hot Jupiter HD189733b, in which enhanced atmospheric escape of the planet was observed 8 hours after a stellar flare \citep{2012A&A...543L...4L}.

\item[(b)] As the planet moves along its orbit, the interaction between a close-in planet with high orbital speeds and the host-star wind can produce bow shocks ahead of planetary motion (\citealt{2010ApJ...722L.168V, 2015ApJ...810...13C}). Because the stellar wind environment along a planet's orbit is not homogenous (due to complex magnetic field topologies of the host star), the strength and geometry of its bow shock change, thus producing spectroscopic transits of varying shapes \citep{2013MNRAS.436.2179L}. Planetary radio aurorae might also change in orbital timescales \citep{2010AJ....140.1929L, 2011MNRAS.414.1573V} -- these radio emissions are pumped by the dissipation of stellar wind power on the magnetosphere of planets, similarly to what is seen in the Solar System \citep{1998JGR...10320159Z}. Therefore, local variations in the stellar wind should affect radio signatures of planetary aurorae in orbital timescales. Finally, notice also that during one orbital revolution, planets might  move in and out of the Alfven surface (e.g., Figure \ref{fig.salfven}), changing  star-planet magnetic coupling (Section \ref{sec.connectivity}) within timescales of days. 

\item[(c \& d)] When magnetic star-planet coupling exists, an exoplanet can excite hot spots on the chromospheres of their  host stars \citep{2015ApJ...815..111S}. This anomalous spot activity depends on the orbital motion of the planet, contrary to regular stellar activity, which is modulated by stellar rotation. Anomalous planet-induced activity observed in CaII K lines has been reported  in some planet-hosting stars \citep{2005ApJ...622.1075S, 2008ApJ...676..628S, 2012PASA...29..141G, 2018AJ....156..262C, 2019NatAs...3.1128C}. The signal however disappears at certain epochs  \citep{2008ApJ...676..628S}.  The on/off nature of such signature can be explained by changes in star-planet coupling, caused by the evolution of stellar magnetic fields, which takes place on timescales of months to decades. Thanks to long-term spectropolarimetric monitoring of stars, complete stellar magnetic cycles are starting to be revealed, and magnetic field evolution identified in several stars \citep[e.g.][]{2013MNRAS.435.1451F, 2018MNRAS.479.5266J, 2022A&A...658A..16B}. Planet-induced radio emission on host stars has been suggested in some recent LOFAR observations  \citep{2020NatAs...4..577V}. Similar to chromospheric anomalous activity, planet-induced radio emission is triggered when the planet orbits within the Alfven surface  \citep{2021MNRAS.504.1511K, 2022MNRAS.514..675K}. Because of stellar field evolution, the visibility of the radio signature can be affected. Furthermore, because the emission is beamed along the surface of a hollow cone, the modulation of this signature is quite complex, requiring a more holistic modelling approach, including orbital motion, visibility of emission cone, stellar magnetic field topology, star-planet magnetic connection, and stellar wind characterisation \citep{2023MNRAS.524.6267K}.  

\item[(e)] The survival of planetary atmospheres depends on the several-Gyr history of the high-energy radiation of the host star \citep{2015A&A...577L...3T, 2019MNRAS.490.3760A, 2021A&A...654L...5P} and also on the evolution of stellar wind particles \citep{2022MNRAS.510.2111K}. Moreover, when substantial material is lost through atmospheric evaporation, the internal structure of the planet readjusts itself \citep[e.g.,][]{2020MNRAS.499...77K, 2021MNRAS.504.2034K}, implying that the evolutionary track of a planet also depends on the history of atmospheric escape. Planetary magnetisation can also evolve over billion of years timescales \citep[e.g.,][]{2013ApJ...770...23Z}, which would alter SPI strengths. Finally, long-term orbital evolution (migration) implies that the incoming stellar radiation and wind properties around the planet also evolve, changing, for example, the atmospheric history of exoplanets \citep{2021A&A...654L...5P}.
\end{enumerate}

%%%%%%%%%%%%%%%%%%%%%%%%%%%%%%%%%%%%%%%
\subsection{The need for contemporaneous multi-wavelength observations to assess SPI} 
In the Solar System, we have a wide variety of measurements that allows us to directly link, for example, how the atmospheres of planets or planetary radio emission respond to variations in the solar wind plasma \citep[e.g.][]{2018NatSR...8..928O, 2023JGRA..12831155L}. In the case of exoplanets, the observational evidence of SPI is more difficult to interpret, one problem being the variability of the signature that can prevent us from reproducing the same findings in follow-up observations \citep[e.g.,][]{2008ApJ...676..628S}. However, if we can identify that the variability of the SPI occurs contemporaneously with variations in stellar magnetism, as recently done by \citet{2018AJ....156..262C}, the very same variability could be the smoking-gun evidence of SPIs.

To interpret (or search for) SPI signatures, I advocate that we need synergetic, contemporaneous observations of the system (e.g., radio plus optical spectropolarimetric data;  chromospheric emission plus spectropolarimetric data). A successful illustration of the importance of coordinating contemporaneous observing campaigns/models came from  the  MOVES collaboration, which aimed at observing the exoplanetary system HD189733 using multi-wavelength observations and different models. Through this collaboration, it was shown that a multi-wavelength approach is necessary to  characterise the variable environment surrounding an exoplanet \citep{2020MNRAS.493..559B}. X-ray  and UV data from HST, XMM-Newton, and Swift were used to model the atmospheric composition of  HD189733b \citep{2021MNRAS.502.6201B}. Additionally, optical spectropolarimetric data \citep{2017MNRAS.471.1246F} were used to characterise the large-scale stellar magnetic field and to drive wind models, which could then be used to assess space weather, radio emission \citep{2019MNRAS.485.4529K} and anomalous chromospheric activity  \citep{2022MNRAS.512.4556S}. 

Although it is possible to coordinate multi-wavelength observing campaigns, coupled to different modelling approaches to characterise SPIs and exo-space weather, in most cases such collaborative efforts do not exist. 
There are different approaches adopted to bypass the lack of contemporaneous data-driven models. One possibility is to create proxy maps of the stellar magnetic field, based on observations of different stars. However, such method should  be used with caution -- even if we can predict the strength of the field using empirical  relations \citep[e.g.,][]{2014MNRAS.441.2361V}, there is no clear trend between the topology of the stellar magnetic field and other stellar properties, such as age and activity. Proxy maps could then lead to substantial differences in the stellar wind properties than those computed using the real reconstructed surface maps. Another possibility is to use magnetic maps of the same star, but reconstructed at  different epochs. However, this method must also be used with caution. In the case of WX UMa, its first magnetic map dates from 2006 \citep{2010MNRAS.407.2269M}, whereas radio observations were conducted nearly 10 years after, between 2014 to 2016 \citep{2021A&A...650L..20D}. Two issues affect the interpretation of radio observations: first, the stellar magnetic field may have evolved during the 10-year interval and, even if that is not the case, the error in the rotation period of the star, when propagated for 10 years, leads to a large uncertainty in the true rotation phase of the star during the radio observations \citep{2022MNRAS.514..675K}.  

 Therefore, when interpreting (and predicting) observables from such signatures, stellar wind models should use magnetic maps that were obtained the closest to the observations as possible.

%%%%%%%%%%%%%%%%%%%%%%%%%%%%%%%%%%%%%%%%%%%%%%%%%%%%%%%%%%%%
\subsection{Multi-fluid models -- going beyond hydrogen/helium}
Currently, the vast majority of models for interpreting observations of evaporation of exoplanetary atmospheres rely on few spectral lines (mostly of hydrogen/helium atoms), preventing us from using the wide wavelength range capabilities of state-of-the-art observatories/instruments that are available. To correctly interpret  observations of  heavier element escape, multi-fluid models, in which each atomic species is described by its own fluid properties, are needed. This will allow the use of observed metal lines as diagnostics of planetary outflow properties, such as mass-loss rates and element compositions. Additionally, with a multi-fluid treatment, one can self-consistently compute fractionation of species and thus, determine how much the hydrogen flow can actually drag other species via collisions. Currently, there are only a few studies investigating flow-flow interactions using a multi-fluid approach in 3D \citep[][see also other works from these teams]{2018MNRAS.481.5315S, 2020MNRAS.491.3435S, 2019ApJ...885...67K}. 

Another point that can be clarified with a multi-fluid formalism is the effect of magnetic fields on  the dynamics and escape of ion and neutral species (see Section \ref{sec.magnetic_escape}). In single-fluid models, it is assumed that neutral-ion collisions are sufficient to couple the ion and neutral species. Because collisions depend on the density of the fluids, this assumption is likely correct for the lower parts of the escaping atmospheres. As one moves to more rarefied conditions, though, this assumption may no longer apply. As charged particles gyrate around magnetic field lines, the neutral and ion flows can respond differently to magnetic fields in the less dense parts of the flow. The distribution of ions under the influence of planetary magnetic fields still needs to be properly evaluated in models, and multi-fluid models can also be of assistance in this area. Additionally, the effects of magnetic fields in charge-exchange reactions are still unknown, as magnetic fields may suppress mixing of stellar and planetary wind fluids, therefore affecting these reactions.

%%%%%%%%%%%%%%%%%%%%%%%%%
\subsection{Time-dependent models -- CME-planetary atmosphere/magnetosphere interactions}
This review focused on interactions mediated by quiescent stellar winds. In addition to a quiescent state, stellar outflows also have bursty phases, when CMEs erupt and propagate through the interplanetary medium. The detection of stellar CMEs is still rare \citep{2022arXiv221105506N}, but it is believed that CMEs could even be the dominant form of mass loss in active stars  \citep{2012ApJ...760....9A,2015ApJ...809...79O, 2017ApJ...840..114C} and might, therefore, be  important  in shaping atmospheric losses of close-in planets orbiting active stars. Studying the impact of flares and CMEs on exoplanets is better performed with time-dependent models. Few studies so far have investigated the impact of CMEs on the magnetospheres and atmospheric structure and escape of close-in exoplanets using 3D models \citep[e.g.,][]{2011ApJ...738..166C, 2017ApJ...846...31C, 2025MNRAS.536.1089H}. 

Currently, one challenge faced by modellers is to realistically describe the conditions of stellar CMEs to be implemented in the simulations, such as CME velocities, densities, magnetisation and overall structure. Models usually adopt measured properties of solar CMEs, but it is not clear whether stellar CMEs would have similar properties as those of solar CMEs (and even then, the diversity of solar CME properties is quite large). Additionally, the frequency of stellar CMEs is poorly constrained, as are stellar CME trajectories. Exoplanets might not interact with every stellar CME, because they may not lie along certain CME trajectories. This is in contrast to stellar winds, which are constantly interacting with exoplanets. CMEs are  likely to be accompanied by associated flares, which generate an increase in high-energy radiation. This increase in incoming radiation  affects the escape rate of planetary atmospheres, as has been studied in 3D models \citep{2019ARep...63...94C, 2022MNRAS.509.5858H}.  Recently, transit observations of the hot Jupiter HD189733b showed an enhancement of  atmospheric evaporation that took place 8h after the host star flared.  \citet{2012A&A...543L...4L} suggested that the violent event and the enhanced evaporation are potentially related to each other. The remaining open question is whether it is the increase in XUV emission caused by the flare or the interaction between an unseen associated CME and the exoplanet (or even both) that caused the increase in the observed evaporation rate \citep{2022MNRAS.509.5858H, 2025MNRAS.536.1089H}. 

%%%%%%%%%%%%%%%%%%%%%%%%%
\subsection{Adjacent fields impacted by SPI studies: planetary habitability and exo-space weather}
In the green box in page \pageref{greenbox_learn}, I highlight important physical aspects one can learn from the study of SPIs. Among the scientific impact of SPI studies that were highlighted were the potential to probe the presence  and constrain planetary magnetic fields, stellar wind environment, atmospheric escape rates and even use radio observations as a planetary discovery technique. Additionally, there are several adjacent fields that the study of SPIs can also impact, such as planetary habitability, exo-space weather and even planet population, as I expand below.

Planetary mass-loss and magnetic fields are relevant for studies of planetary habitability \citep[e.g.][]{2007AsBio...7..185L,2007AsBio...7...85S}. Due to geometrical effects, close-in planets interact much more frequently with stellar CMEs than planets at larger orbits. Certain types of stars, such as M dwarfs, the prime targets for detecting terrestrial planets in the habitable zone, remain active for a long part of their lives \citep{2008AJ....135..785W} and have high flare and CME rates \citep{2016A&A...590A..11V}. SPI studies can quantify the life-integrated amount of  mass lost by potentially-habitable planets orbiting M dwarfs, similar to systems such as TRAPPIST-1 and Proxima Cen, which can then inform planetary habitability studies. 

In addition to possibly influencing planetary habitability, when exoplanets lose mass, their orbital evolution changes, which may lead to planetary engulfment \citep{2009ApJ...705L..81V, 2016A&A...591A..45P}. Exoplanetary mass loss also regulates angular momentum evolution \citep{2014ApJ...788..161T}, which is deeply connected to magnetic field generation \citep{2012Icar..217...88Z}, which in turn may affect atmospheric retention \citep{2015ApJ...813...50K, 2021MNRAS.508.6001C}. Extreme mass loss can remove planetary atmospheres turning planets that once were gas giants into naked cores, reducing therefore their lifetimes  \citep{2003ApJ...598L.121L, 2004A&A...419L..13B,  2004A&A...418L...1L, 2013MNRAS.433.3239K}. As a consequence, evolution of planetary mass loss is also crucial for understanding exoplanet populations \citep{2009MNRAS.396.1012D, 2021MNRAS.504.2034K} and the structure of any given planet \citep{2014ApJ...795...65J}. SPI studies can reveal new trends (e.g., a relation between evaporation rate and stellar wind power) that could inform future studies of planet population synthesis.

Finally, understanding how the Sun and other stars create space weather is a key theme in the current research landscape \citep[e.g.,][]{decadal}. By applying stellar wind and CME modelling to spectroscopic transit data, we can quantify the space weather around exoplanets, such as velocity, density, and  mass-loss rates. Stellar winds carry away angular momentum, and they thus regulate stellar rotation \citep{2021LRSP...18....3V}. The latter affects the dynamo operating in stellar interior, which produces magnetism that is observed at the stellar surface. 

%%%%%%%%%%%%%%%%%%%%%%%%%%%%%%%%%%%%%%%%%%%%%%%%%%%%%%%%%%%%%%%%%
\section{CONCLUSIONS}

The scope of this review article was to present an overview of 3D MHD numerical models developed in the study of SPIs. I covered here three main interaction types, all of which are in one way or another mediated by stellar magnetism. (1) Radiative interactions that can give rise to an expansion of exoplanetary atmospheres and their subsequent escape. (2) Particle interactions (or flow-flow interactions mediated by the stellar wind) that generate different morphological features that can be captured in observations. (3) Magnetic interactions that allow for direct magnetic connection between stellar and planetary field lines. The strength of the SPI depends on the architecture of planetary systems, as well as on the age and activity level of the host stars. In particular, exoplanets in close-in orbits and/or orbiting active host stars can undergo strong  physical interactions. Comparatively, some of these interactions are negligible or absent in the present-day Solar System. The key points discussed here and future prospects for the field are listed next.

\begin{summary}[SUMMARY POINTS] 
\begin{enumerate} 
\item Models are fundamental to interpret and  guide  observations. The powerful combination of observations and models allows us to extract important physical parameters of the system, such as, planetary magnetic fields, stellar wind properties, etc (see box in page \pageref{greenbox_learn}).
\item Models with simplified symmetries (e.g., spherical 1D models) are excellent to treat processes that are more numerically expensive, such as chemistry of outflowing atmospheres. However, one drawback is that these models cannot capture  the  truly 3D asymmetric structure of the upper atmospheres/magnetospheres of planets. These upper regions are directly shaped by interactions with the stellar wind and their dynamics affected by the presence of a planetary magnetic field. This is an important limitation that can only be overcome with multi-dimensional models. 
\item Star-planet interactions generate spatially asymmetric features (e.g., planetary material trailing the orbit, shock formation), thus requiring the use of three-dimensional models. 
\item Star-planet interactions  vary in different timescales (from hours to giga-years) that are related to both planetary  (orbital motion, rotation) and stellar  (flares, cycles, and long-term evolution) properties. Understanding these variations may require time-dependent models. 
\end{enumerate} 
\end{summary}

\begin{issues}[FUTURE ISSUES] 
\begin{enumerate}
\item Observations conducted at different wavelengths and simultaneously can help reduce controversial interpretations of SPIs. Ideally, future 3D models should be informed by (near-)simultaneous, multi-wavelength observations. The use of observations is twofold: some generate inputs for models (e.g., stellar magnetic field maps), whereas others are fitted by models (e.g., spectroscopic transits). Taking observations and models together, we have a powerful tool to derive physical properties of the system that would otherwise remain unknown. 
\item As we start to model planetary atmospheric escape of species heavier than hydrogen/helium, there  is the need of  expanding the efforts of developing 3D multi-fluid simulations. With multi-fluid treatments, one can self-consistently compute fractionation of species and thus, determine how much the hydrogen flow can actually drag other species via collisions. Multi-fluid models are also important for better understanding the effects of planetary magnetic fields on the dynamics and escape of ion and neutral species, and modelling charge-exchange between stellar wind and planetary winds, as the physics of charge exchange might not be well captured in single-fluid simulations.
\item Star-planet interactions  can operate in different timescales from hours/days to giga-years. In particular, the effects of bursty events such as flares and CMEs on the planetary magnetosphere and atmosphere is more appropriately studied with time-dependent models. A particular difficulty in this aspect is to correctly capture the physical conditions of stellar CMEs, as at the moment they are not well constrained.
\item Spectroscopic transit observations are powerful to detect atmospheric escape and probe the interaction with stellar outflows. However,  models have shown that different combinations of stellar wind properties and planetary magnetisation can lead to similar synthetic transit Ly-$\alpha$ line profiles. One potential way to break the degeneracy between stellar wind properties and planetary magnetisation is to use  multiple spectral lines that diagnose escape at different altitudes. For example, the Ly-$\alpha$ line wings probe the region of interaction between the planetary upper atmosphere and the stellar wind, whereas heavier species are more likely to probe regions that are less affected by the stellar wind interaction and more affected by the confinement due to the planet's magnetic field. 
\end{enumerate}
\end{issues}

\section*{DISCLOSURE STATEMENT}
The author is not aware of any affiliations, memberships, funding, or financial holdings that might be perceived as affecting the objectivity of this review.

\section*{ACKNOWLEDGMENTS}
I acknowledge funding from the European Research Council (ERC) under the European Union's Horizon 2020 research and innovation programme (grant agreement No 817540, ASTROFLOW) and from the Dutch Research Council (NWO) under research programme ``NWO Talent Programme VICI 2023" (project number VI.C.232.041). This work used the Dutch national e-infrastructure with the support of the SURF Cooperative using grant nos. EINF-2218 and EINF-5173. I also thank the reviewer and scientific editor of this chapter, Prof.~Sarah Gibson, for her comments and suggestions as I was writing this review. I also thank the following colleagues for discussions and providing feedback in earlier versions of this manuscript: Dr.~Carolina Villarreal D'Angelo, Dr.~Salvatore Colombo, Dr.~Robert Kavanagh, Dr.~Maxim Khodachenko, and Dr.~Ildar Shaikhislamov.

\newpage
\setcounter{figure}{0}
\renewcommand\thefigure{Supplemental \arabic{figure}}  

\section*{Supplemental Material to ``Star--Planet Interactions: A Computational View''}
\begin{figure}[!h]
\includegraphics[width=1\textwidth]{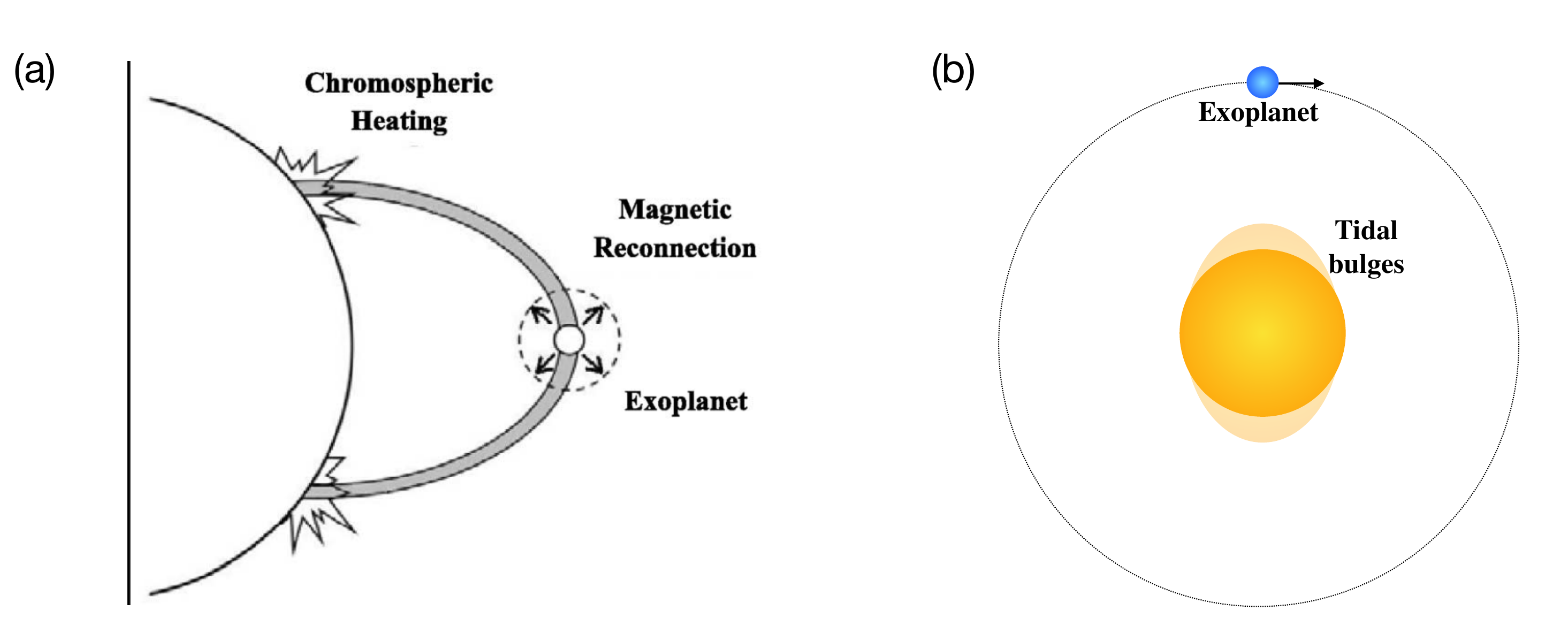}
\caption{Earlier suggestions of SPI evolved from the analogy with binary stellar systems. (a) Magnetic field lines of two magnetised bodies could reconnect, accelerating particles that would impact the star. Panel reproduced with permission from Ip et al (2004); copyright AAS.
 (b) Two bodies in close orbit could generate two expanding/contracting  tidal bulges that would trigger waves and non-radiative energy.  Both processes (a) and (b) would generate localised heating in the form of hotspots in the outer stellar atmosphere layers, mimicking the effects of stellar activity. 
\label{fig.earlier_SPIs} }
\end{figure}
\begin{figure}[!h]
	\centering	\includegraphics[width=.7\textwidth]{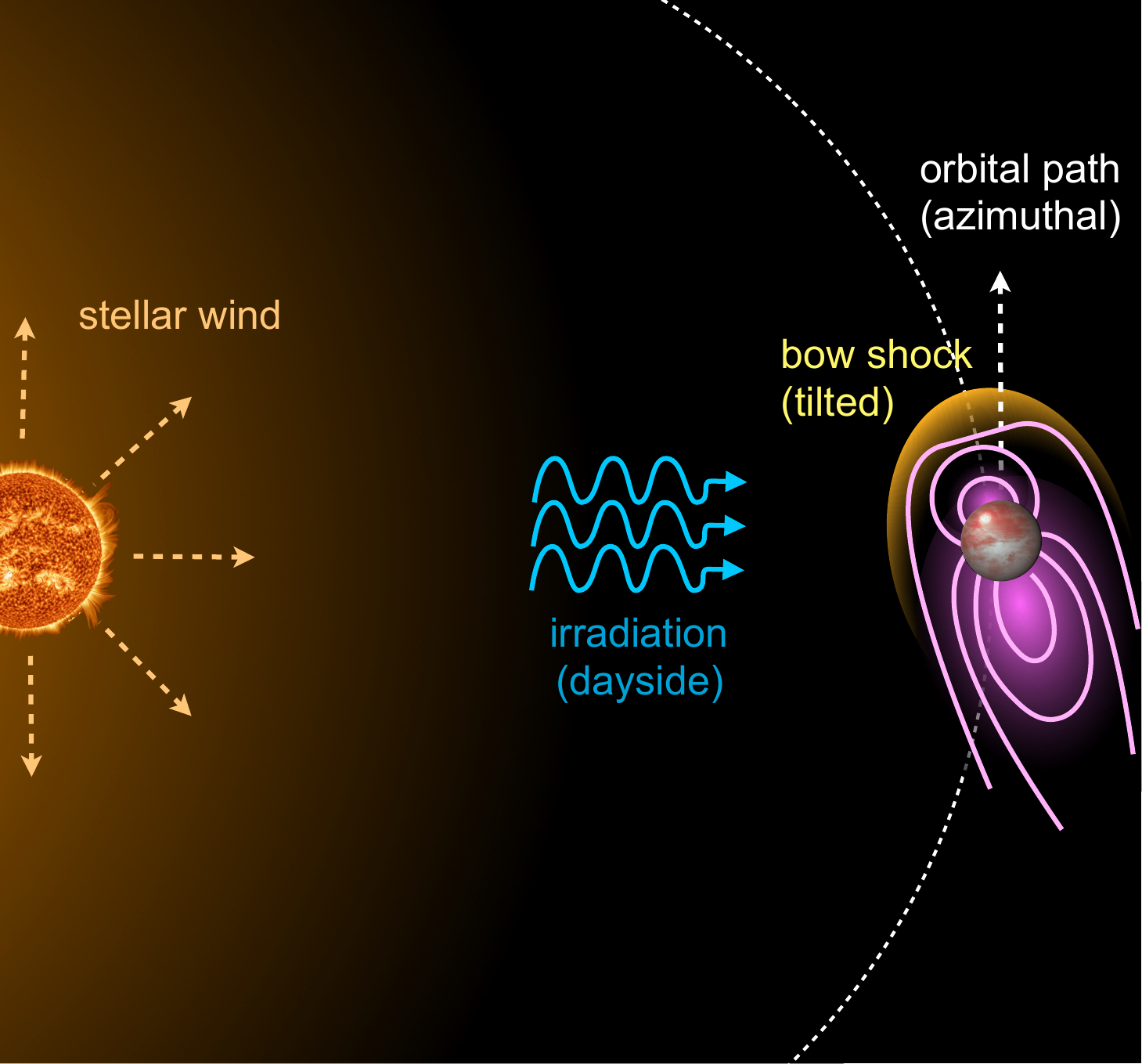}
	\caption{Due to high orbital speeds of close-in planets, bow shocks are formed at an angle with the orbital motion, while stellar irradiation always impact on the dayside of the planet.}\label{fig.mismatch}
\end{figure}

\end{document}